\documentclass[utf8]{frontiersFPHY} 
\setcitestyle{square} 
\usepackage{url,hyperref,lineno,microtype,subcaption}
\usepackage[onehalfspacing]{setspace}
\usepackage{ulem}

\newcommand{\halpha}{H\ensuremath{\alpha}}

\newcommand{\arcsec}{\ensuremath{^{\prime\prime}}}
\newcommand{\unit}[1]{\ensuremath{\mathrm{\, #1}}}
\newcommand{\figref}[1]{Figure \ref{#1}}

\newcommand{\code}[1]{\ensuremath{\textsf{#1}}}

\newcommand{\fnc}[1]{\code{#1}}

\newcommand{\secref}[1]{$\S$\ref{#1}}
\newcommand{\linew}{\ensuremath{\delta\lambda}}
\newcommand{\mhz}{\ensuremath{\unit{mHz}}}
\newcommand{\pmode}{p-mode}

\def\keyFont{\fontsize{8}{11}\helveticabold }

\def\firstAuthorLast{Tarr {et~al.}} 
\def\Authors{Lucas A. Tarr\,$^{1,*}$, Adam R. Kobelski\,$^{2}$, Sarah A. Jaeggli\,$^{1}$, Momchil Molnar\,$^{3, 4}$, Gianna Cauzzi\,$^{3,5}$,Kevin P. Reardon\,$^{3,4}$}

\def\Address{$^{1}$ National Solar Observatory, Makawao, HI, USA \\
  $^{2}$ NASA, MSFC, Huntsville, AL, USA\\
  $^{3}$ National Solar Observatory, Boulder, CO, USA\\
  $^{4}$ Department of Astrophysical and Planetary Sciences, LASP, University of Colorado, Boulder, CO, USA\\
  $^{5}$ INAF–Osservatorio Astrofisico di Arcetri, Largo Enrico Fermi 5, 50125 Firenze, Italy}
 \def\corrAuthor{Lucas A. Tarr, 22 Ohia Ku St, Makawao, HI, 96768}
 \def\corrEmail{ltarr@nso.edu}

\begin{document}
\onecolumn
\firstpage{1}

\title[H-alpha versus ALMA brightness temperature]{Spatio-Temporal Comparisons of the Hydrogen-Alpha Line Width and ALMA 3 mm Brightness Temperature in the Weak Solar Network} 

\author[\firstAuthorLast ]{\Authors} 
\address{} 
\correspondance{} 

\extraAuth{}

\maketitle

\begin{abstract}

Comparisons between the Atacama Large Millimeter/sub-millimeter Array (ALMA) 3 mm emission and a range of optical and UV solar observations have found the strongest correspondence between the width of the hydrogen alpha line at 656.3 nm and the 3 mm brightness temperature.  
Previous studies on the oscillatory power of \pmode{}s using ALMA Band 3 and Band 6 data in the 3 to 5 minute period bandpass have found a confusing mix of results, with many reporting a complete lack of the \pmode{} enhancement typically found in other chromospheric observables.
We study these issues using an extensive, publicly available coordinated data set targeting a region of weak network flux near disk center at time SOL2017-03-17T15:42-16:45. 
We focus on the Interferometric Bidimensional Spectropolarimeter (IBIS) H-alpha and ALMA 3 mm data series.

We confirm the strong correlation between the H-alpha line width and the 3 mm brightness temperature, but find a bimodal relation between the two diagnostics, with a shallower slope of 7.4e-5 \AA{}/K in cooler regions and steeper slope of 1.2e-4 \AA{}/K in hotter regions.
The origin of the bimodal distribution is unknown, but does hold for the duration of the observations.
Both slopes are steeper than a previously reported value, but this is likely due to systematic differences in the analysis.
We then calculate the oscillatory power in the  H-alpha and 3 mm data.  
The IBIS data clearly show the \pmode{} oscillations in spatially averaged power spectra while the ALMA data do not.
However, when we remove IBIS data at times corresponding to the ALMA calibration windows, the spatially averaged power spectra for the two data series are nearly identical, with a Pearson correlation coefficient of $0.9895$.
Further, the power in the two bands remains strongly correlated when the spatial information is retained but the power is integrated over different temporal frequency bands.  
We therefore argue that the lack of observed \pmode{}s in the ALMA data may be predominantly due to spectral windowing induced by the timing and duration of the calibration observations.
Finally, we find that spatial maps of oscillatory power at 3 mm display the pattern of magnetic shadows and halos typically displayed by other chromospheric diagnostics.

\tiny
 \keyFont{ \section{Keywords:} Solar Physics, Solar chromosphere, Solar oscillations, Solar radio emission, Solar activity, Solar Magnetic Fields}
\end{abstract}

\section{Introduction} \label{sec:intro}

The solar chromosphere is a richly structured and dynamic plasma environment typified by increasing temperature, decreasing density, and the rapid transition of numerous plasma properties with height, such as ionization fraction (towards more highly ionized states) or the ratio of thermal to magnetic energy (towards magnetic dominance) \citep{Bray:1974, Leake:2014}.
The chromosphere acts as a conduit for energy to propagate from the mechanical reservoir of the upper convection zone into the corona where it can heat the coronal plasma, accelerate the solar wind, and power flares and eruptions.
Determining how the chromosphere is structured is a critical task towards identifying which energy transfer mechanisms are dominant in each type of solar environment---quiet Sun, plage, within coronal holes, at the center of sunspots, etc.---and how they evolve over time.
Given the rapid transitions within the chromosphere, multiple radiative diagnostics that are sensitive to different properties of the plasma must be observed at the same time and understood as a whole in order to reconstruct the chromospheric structure.

A longstanding question in solar physics is the nature of chromospheric oscillations and their relative importance for the propagation of energy between the photosphere and the corona (see review by \citet{Khomenko:2013} and references therein).
Observations at millimeter wavelengths provide a rather direct measurement of the chromospheric electron temperature and are therefore an important diagnostic of these oscillations \citep{Loukitcheva:2006} especially when obtained with the good spatial and temporal resolution provided by the Atacama Large Millimeter/sub-millimeter Array \citep[ALMA; ][]{Wooten:2009, Hills:2010}.
However, reports of oscillations in the ALMA data sets have so far been rather mixed.

\citet{Jafarzadeh:2021} presented results from a wide variety of solar data sets and features observed with Band 3 and Band 6 data during the ALMA Cycle 4 solar campaign.  
They found evidence what is expected to be nearly ubiquitous enhancement in oscillatory power in 3-5 \mhz{} \pmode{} range typically found in chromospheric diagnostics (5 and 3 min periodicity, respectively) in just two of the ten data sets they analyzed.
The two sets of observations that did contain evidence of chromospheric oscillations were studied in more detail in \citet{Nindos:2021}, who did detect \pmode{} power in both Band 3 and Band 6 ALMA data.

\citet{Patsourakos:2020} found oscillatory power at \pmode{} frequencies in their Band 3 quiet Sun observations at multiple heliographic angles in a center-to-limb study.
However, when looking for spatial coherence in the oscillatory power, they found no coherence at scales above the ALMA spatial resolution, and reported no correlation with the underlying pattern of either brightness structure or network/internetwork regions.
They also reported that frequency-integrated maps of the power spectra in select bands did not reveal the power-shadow or -halo features typically found in photospheric and chromospheric diagnostics \citep{Brown:1992,Braun:1992,Vecchio:2007}.

\citet{Narang:2022} calculated the oscillatory power in a region of plage using ALMA Band 6 observations, which correspond to lower heights and cooler temperatures relative to Band 3, although there can be significant overlap between the contribution functions for the two channels \citep{Wedemeyer:2016}.
They investigated the spatially averaged power over the ALMA field of view as well as the spatial distribution of frequency-integrated power in different temporal frequency ranges. 
Similar to most of the data series analyzed by \citet{Jafarzadeh:2021}, these authors did not find significant power in the \pmode{} band of 3-5 minute periods.
They also did not find any strong correlations between the spatial distribution of oscillatory power in the ALMA data compared to transition region and coronal data sets from the Interface Region Imaging Spectrograph (IRIS) and the Atmospheric Imaging Assembly (AIA) on the Solar Dynamics Observatory (SDO).

\citet{Molnar:2021} reported somewhat similar results to the above for power spectra of the Band 3 observations of a region of network flux near an active region.
Their plots of the spatially averaged power spectra (e.g., their Figure 14) do not show any prominent peak at \pmode{} frequencies.  
They do, however, report fairly strong spatial variations in the power spectra that correlate with different temperature regimes, and all have the same power law trend.
They also find several power enhancements above the power law trend in specific bands at higher frequencies, from 10 to 50 \mhz{}.

\citet{Chai:2022} found very clear indications of power at 5 \mhz{} in a sunspot umbra with their ALMA Band 3 observations.
These oscillations showed good correspondence with simultaneous velocity proxies in the \halpha{} line.
At higher frequencies, umbral and penumbral regions had essential the same spectrum. 
Their quiet Sun regions did not show any prominent enhancement in the 3-5 \mhz{} range, but did show some excess power around 30 \mhz{}, perhaps similar to the findings of \citet{Molnar:2021}.

\citet{Nindos:2021} included an appendix considering the effect of observation duration and calibration windows on the measured power spectrum. 
Using a monochromatic source as an example, they found that the windowing effect for their ALMA cycle 4 observations did not reduce the spectral width of the recovered source (which is due to finite frequency resolution), but did reduce the peak power as well as modify the side-lobe size and pattern.
We will use a multi-instrument data set to study the effect temporal sampling and calibration windows of the computed power spectrum in more detail.

We present here analysis of a combined data set described in \citet{Kobelski:2022} that includes 3 mm (or 100 GHz; Band 3) interferometric observations from the ALMA, sensitive to the chromospheric temperature, and spectral-imaging observations of the hydrogen \halpha{} line at 656.3 nm from the Interferometric Bidimensional Spectropolarimeter \citep[IBIS; ][]{Cavallini:2006} at the Dunn Solar Telescope.
The combined data set covers $\sim 60 \unit{min}$ of observations of a small, bipolar region of network flux located near disk center on 2017-03-21.
The millimeter radiation provides a direct measure of the electron temperature in the chromosphere, but exactly where the emission forms, how it varies based on dynamic phenomena or on the surrounding magnetic structure, and how it relates to other diagnostics is still under investigation \citep{Wedemeyer:2020,Alissandrakis:2020,Narang:2022}.

Our aim in this paper is to explore spatial and dynamic correlations in our observations of network magnetic fields.
\citet{Molnar:2019} found a strong positive correlation between the ALMA 3 mm brightness temperature and the \halpha{} line width in an area of network flux near an active region.
A linear fit to the trend gave a slope of $6.12\times10^{-5}$\AA{}$/\unit{K}$.
\citet{Kobelski:2022} confirmed this strong correlation between the two quantities for a region of weakly enhanced network magnetic field in the quiet Sun.
They reported a steeper slope of $1.1\times10^{-4}$\AA{}$/\unit{K}$ over a narrower range of observed temperatures, due to their more quiet target region.
They also used a slightly different definition of the line width compared to \citet{Molnar:2019}, which affects the value of slope but not the goodness of fit.
In any case, the correspondence between the 3 mm brightness temperature and the width of the \halpha{} line is striking, to the extent that the two observables, when sampled at the same spatial scale, cannot readily be distinguished. 

By comparing the power spectra derived from the \halpha{} line width to those derived from the 3 mm data, we argue in this paper that the lack of detected \pmode{}s in the ALMA data in previous studies may be an effect of the windowing due to the timing of calibration data acquisition. 

In Section \ref{sec:obs} we describe the observations used in the analysis.
In Section \ref{sec:dataprep} we describe how we prepped the original data, specifically how we characterised the \halpha{} data for use in later analyses.
In Section \ref{sec:joint} we discuss properties of the joint distribution of the \halpha{} and 3 mm data series.
In Section \ref{sec:time} we discuss aspects of each time series of data in terms of the average power spectra and spatial distribution of oscillatory power.
We briefly discuss the results in Section \ref{sec:discussion} and finally conclude in Section \ref{sec:conclusion}.

\section{Observations}\label{sec:obs}

\begin{figure}[ht]
    \centering
    \includegraphics[width=\textwidth]{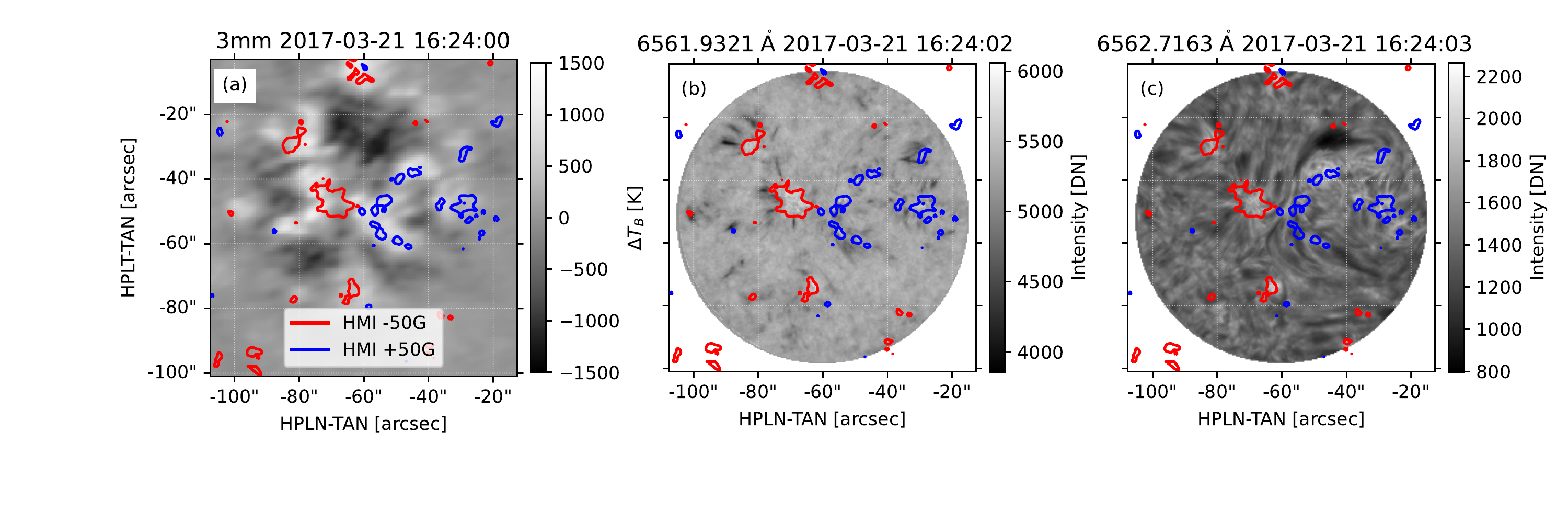}
    \caption{(a) ALMA Band 3 brightness temperature relative to the mean temperature, (b) intensity in the IBIS \halpha{} blue wing at 656.19 nm, and (c) intensity near the \halpha{} line core at 656.27 nm.  In all three panels the red and blue contours mark the $\pm50$ Gauss line of sight magnetic field measured by HMI.}
    \label{fig:overview}
\end{figure}

The present work uses two data series taken from the more extensive coordinated data set described in \citet{Kobelski:2022}\footnote{In keeping with their terminology, we use \textit{data series} to refer to data from a single instrument and \textit{data set} to refer to the entire collection of coaligned data series from all instruments}.  
The full data set includes coordinated observations between ALMA and multiple instruments at each of the  Dunn Solar Telescope (DST), IRIS, Hinode spacecraft, and SDO facilites; the Nuclear Spectroscopic Telescope Array (NuSTAR) also observed the same solar target within several hours of our observations.  
The calibration of each data series, the coalignment of all series to each other, and the solar target are fully described in that paper, so here we only provide a short overview of the details relevant to the present work.
\citet{Kobelski:2022} have made the full data set publicly available at \url{https://share.nso.edu/shared/dkist/ltarr/kolsch/}.

ALMA was operated in configuration C43-1 and observations were taken using Band 3, with a central frequency of 100 GHz (3 mm wavelength) and a bandwidth of 18 GHz.
With this configuration and frequency, ALMA's spatial resolution is approximately $3\arcsec$ on the sky, though the beam shape is elliptic.
The data series was self-calibrated using the \fnc{CLEAN} algorithm and produced output at a 2 second cadence, or 1515 spatial maps of the relative brightness temperature $\Delta T_B$ over the entire time series.
The time series includes five $\approx 10$ minute acquisitions, each separated by 3 minutes of calibration data. 
Given the quiet Sun target and its proximity to disk center, we use the mean temperature of 7300 K reported in \citet{White:2017} for quiet Sun disk center ALMA 100 GHz observations to define the zero point of these observations, i.e., $T_B = \Delta T_B + 7300\unit{K}$.
The details of the ALMA data set are described further in \citet{Kobelski:2022}.
The spatial maps show features over approximately a 60\arcsec{} diameter field of view, see \figref{fig:overview}(a).
We note, however, that the spatial reconstruction of this interferometric data does not produce a hard cutoff in the field of view. 
Comparison to the \halpha{} data carried out in \secref{sec:dataprep} does show well-correlated variations out to the edge of the IBIS field of view, which does have a hard cutoff due to the circular field stop.
With this in mind, care must be taken when comparing the two data series.

The dual Fabry-P\'erot based IBIS instrument was used to scan the \halpha{} line at 656.3 nm repeatedly, executing 1800 scans during the observations.
Each scan consisted of 26 unequally spaced wavelength positions, using tighter sampling in the line core of $\approx12.5\unit{pm}$ and $\approx19.1\unit{pm}$ in the line wing, and covering a total of $\pm2$ \AA{} about line center.  
At each wavelength scan step, IBIS imaged a 2D spatial field of view 90\arcsec{} in diameter. 
Each step in the spectral scan lasted $0.167\unit{s}$, giving a total cadence of $4.3\unit{s}$ for a single complete scan of the spectral line; i.e., all 26 spectral points over the spatial field of view.

Our primary focus is on the $\approx60\unit{min}$ of observations with Band 3 of ALMA, which has a central frequency of 100 GHz or 3 mm in wavelength, and the spatially- and temporally-overlapping imaging-spectroscopic observations of the $656.3\unit{nm}$ \halpha{} line that last $\approx 120\unit{min}$ in total and completely overlap the ALMA observations.
The target was a small, bipolar patch of enhanced network magnetic flux approximately $(-60\arcsec,-50\arcsec)$ from disk center, observed on March 21 2017.
\figref{fig:overview} presents an overview of the region and our observations.
Panel (a) shows the ALMA $\Delta T_B$.
Panels (b) and (c) show intensity maps from two of the twenty six positions of the IBIS spectral scan, in the blue wing (656.19 nm) and near the center (656.27 nm) of the \halpha{} line, respectively.
The red and blue lines in every panel respectively mark the $-50$ and $+50$ Gauss contours of the line of sight magnetic field measured by HMI and pulled from the \fnc{hmi.M\_720s} data series.
The dark filament that can be seen in the \halpha{} core lies along the polarity inversion line of the small bipole near the center of the IBIS field of view.
The filament varies in extent and intensity, persists throughout the observation, and occasionally shows very strong signals in the blue wing around -0.8 \AA{} from the line core.
Those dynamics will be considered in a separate work.

The magnetic flux in this area was likely associated with the decay of NOAA Active Region 12639, although by the time of our observations it had essentially merged with the background network.
In fact, it had decayed substantially since being observed by another group in the ALMA Cycle-4 campaign \citep{Shimizu:2021} two days prior to our own observations, a happy coincidence that could be exploited in future studies.
While decaying, \citet{Kuhar:2018} were able to detect microflaring from this particular region a few hours after our observing window using X-ray data from NuSTAR with a total thermal energy of order a few $\times10^{26}\unit{erg}$.
Given their complex spatial and temporal morphology, and the ratio of thermal to nonthermal emission allowed by NuSTAR's sensitivities, these authors conclude that the impulsive events in this region classify as (small) quiet sun microflares, but not the elementary heating events proposed by \citet{Parker:1988}.

For use in later analyses we define the set of ALMA times as $\{t_A\}$ and the set of IBIS times as $\{t_I\}$, where times are defined at the center of the integration for ALMA and at the center of each spectral scan for IBIS.
These sets exclude ``bad'' data frames during the calibration windows (ALMA) and periods of bad seeing (IBIS).
We then define their set intersection as $\{t_O\}=\{t_A\}\bigcap\{t_I\}$, where the new, ``overlapping'' set is defined as the set of all closest-in-time pairs (one ALMA, one IBIS) with no repeated times from either data series.

To elaborate further, the ALMA data have a nominal cadence of 2 sec, but approximately every ten minutes it has an approximately 3 minute data gap due to the calibration window.  
The IBIS data have a nominal cadence of approximately 4.3 sec, but have occasional spans of bad data that we do not use due to poor seeing.  
We wish to pair up elements from each data series.  
The goal of this task is not to temporally co-align the two data series (for that purpose we would simply interpolate the data), but instead to generate data series with similar characteristics in terms of duration and sampling, from which we may calculate the temporal power spectrum of each series and understand the effect of sampling on these calculated spectra by comparing them to the power spectra calculated from each of the full time series.  
To achieve the best statistics, we wish to use the maximum number of samples from each data series without using any sample twice.  
In a very rough sense, this means that every other sample from the ALMA time series will be paired with one IBIS sample. 
But because the two cadences are not related by an integer multiple, and because there are (independent) data gaps present in each data series, the actual pairing is more erratic.
Given the ratio in sampling between the two data series, roughly every 8-10 IBIS timesteps an additional ALMA timestep is skipped, but this is of course modified any time there is missing data from either time series.

The final result of the set intersection produces 685 pairs of samples from the ALMA and IBIS time series, with a (very roughly) 4 second cadence.
We do not enforce a maximum allowed time difference between any pair of elements in $\{t_O\}$.  The largest temporal offset between any pair of ALMA-IBIS samples is 1.7s, 8 pairs have $\Delta t >1s$, and the remaining 677 pairs are essentially evenly distributed between $0<\Delta t<1s$ in $0.2s$ bins.

\section[H-alpha line width]{\halpha{} line width measurements}\label{sec:dataprep}

\begin{figure}[ht]
    \centering
    \includegraphics[width=0.5\textwidth]{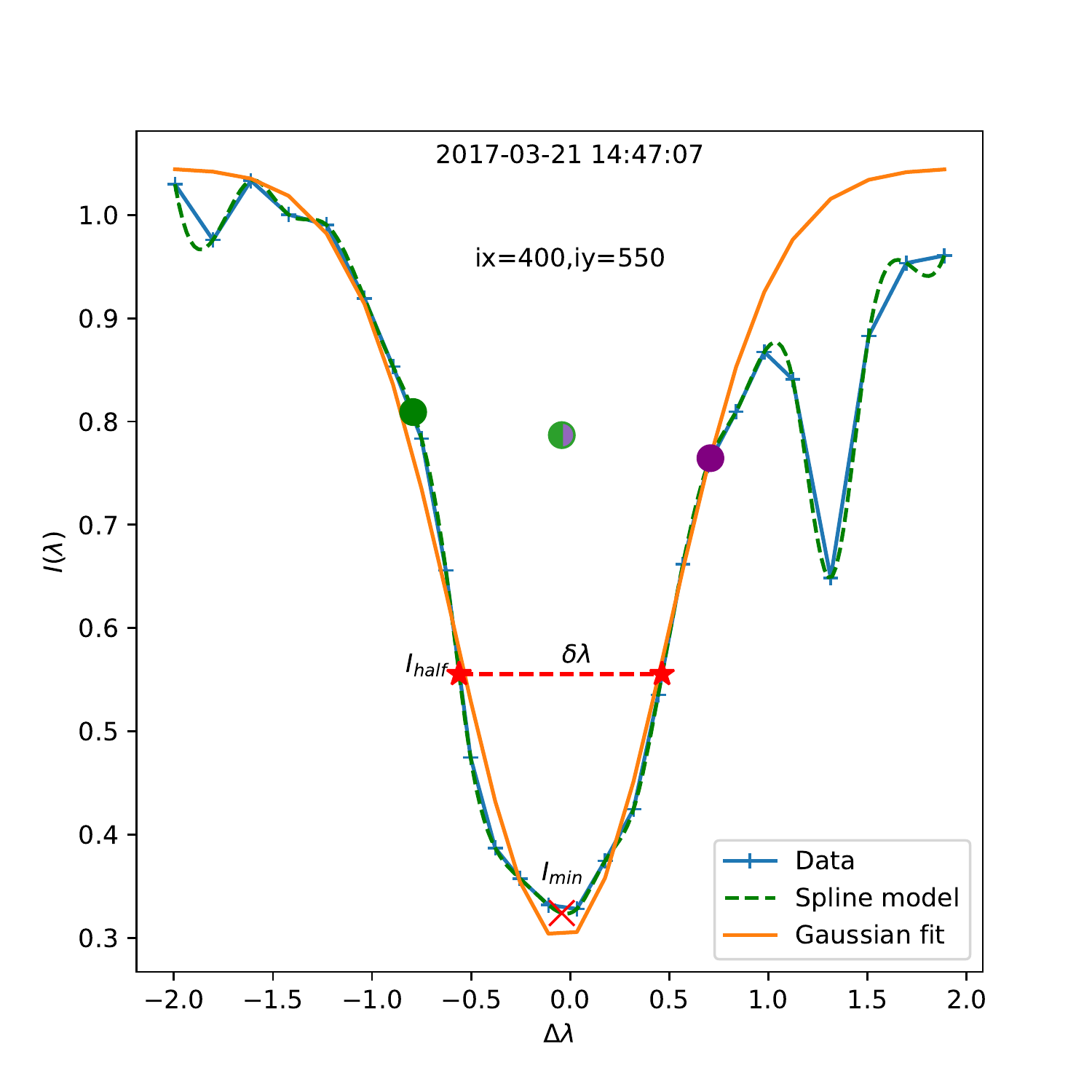}
    \caption{Example calculation of the \halpha{} line width. The x-axis is wavelength in \AA~from the nominal line core, while the y-axis is the profile intensity, normalized to the maximum of the average profile over the whole sample.  This figure is reproduced from \citet{Kobelski:2022} by permission of the AAS.}
    \label{fig:linewidthexample}
\end{figure}
Following in the footsteps of \citet{Molnar:2019} who found the remarkable relation between the width of the \halpha{} line and the 3 mm brightness temperature, we begin by characterizing the \halpha{} line in terms of the minimum intensity, center wavelength (i.e., Doppler velocity), and spectral width.

We characterize the line independently at each spatial location and each time in the data series using the method described in \citet{Kobelski:2022}$\S4.2$, which in turn is based on \citet{Cauzzi:2009}.
\figref{fig:linewidthexample} shows the various steps in the process where the final result, the line width, is indicated by the red dashed line.
The method is, in effect, a version of the full width at half max, but modified to isolate only the chromospheric portion of the line profile around the line core.
First, we construct a spline-interpolated model (green dashed line) of the measured line profile (blue `+' symbols and straight lines).
We define the center wavelength as the center of the best-fit Gaussian (orange) to the central 12 measured points.
The intensity minimum (red cross; $I_\text{min}$) is the value of the spline model at the center wavelength.
Next we define an upper threshold as the average intensity at $\pm0.75$ \AA{} from the line center (marked by the green and purple dots) using the spline model.
The intensity midpoint between the upper threshold and the minimum is marked in each wing of the line by the red stars.
The final width $\delta\lambda$ is the distance between them calculated using the spline model, shown as the red dashed line.

We use $\pm0.75$ \AA{} to define the upper intensity threshold in order to avoid the influence of a telluric H$_2$O line at $+1.5$ \AA{} relative to \halpha{}.
In contrast, \citet{Molnar:2019} used $\pm1.0$ \AA{} as their upper bound, as their data were obtained under drier atmospheric conditions at the DST and do not appear to suffer from telluric H$_2$O contamination. 
The clean signal they obtained throughout the entire line allowed those authors to vary the parameters defining the line width and test the resulting correlation between the width and the ALMA brightness temperature.
They found that, while the slope of a linear fit between $\delta\lambda$ and $T_B$ varied depending the parameters, the correlation coefficient was relatively similar.
We address this point further in the discussion in \secref{sec:discussion}.

\begin{figure}
    \centering
    \includegraphics[width=\textwidth]{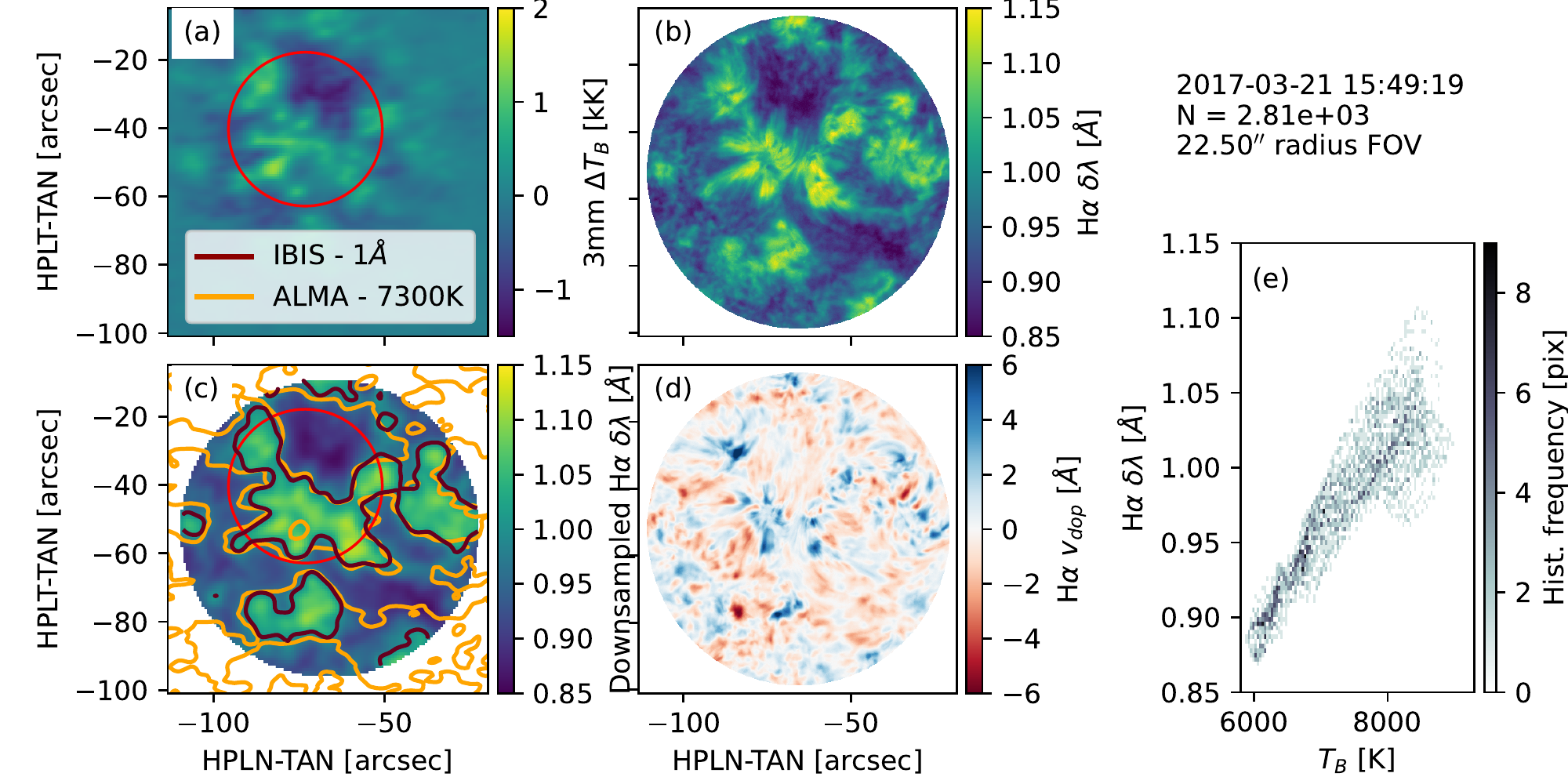}
    \caption{Overview of the primary data series used for this study: (a) ALMA Band 3 relative brightness temperature $\Delta T_B$; (b) full resolution IBIS \halpha{} line width; (c) convolved and spatially down-sampled line width; (d) full resolution \halpha{} Doppler velocity; and (e) the joint probability distribution of brightness temperature and the line width within the 45\arcsec{} diameter red circle, which is a conservative estimate of the overlapping FOV.  Panel (c) shows contours of the ALMA 7300 K level (orange) and IBIS 1 \AA{} line width level (dark red).  An animation based on panels (a), (b), (c), and (d) is included in the online material.}
    \label{fig:tblwpddshist}
\end{figure}

\figref{fig:tblwpddshist} presents an overview of the data analyzed at a single time, 15:42:12 UT.
In the top row we plot (a) a spatial map of the ALMA Band 3 relative brightness temperature $\Delta T_B$ and (b) the calculated IBIS \halpha{} line width \linew{} at full IBIS resolution.
In panel (c) we show a map of the line width after being convolved with a 2D Gaussian (described next) and interpolated to the locations of the ALMA pixel centers. 
Panel (d) shows the line-of-sight Doppler velocity of the \halpha{} line at full IBIS resolution.  
Contours of both data sets are shown in Panel (c) (see labels in Panel (a)) and provide a by-eye demonstration of the good correlation between $\Delta T_B$ and \linew{}.
An animation of these four panels for the 685 time steps in $t_O$ is available in the online material.
Panel (e) shows the joint distribution of $T_B$ and \linew{} calculated using only pixels within the red circle from panels (a) and (c), or 2813 pixels for this single time.
This $45\arcsec$ diameter circle marks the conservatively-defined overlapping FOV for the two instruments, as described below.

We now discuss our methodology for spatially smoothing and interpolating the IBIS data to best match the ALMA data.
In general, the ALMA spatial resolution pattern is returned by the \fnc{CLEAN} algorithm as a best-fit ellipse to the primary beam.
An equivalent circular Gaussian resolution element can be defined by $\sqrt{b_\text{min}b_\text{max}}$ where $b_\text{min}$ and $b_\text{max}$ are the major and minor radii of the ALMA beam projected onto the sky.
The beam shape changes over the course of our observations at the $2\%$ level, with maximum, minimum, and median values of the Gaussian FWHM of 3.28\arcsec{}, 3.21\arcsec{}, and 3.236\arcsec{}, respectively.

Instead of using the reported ALMA beam parameters we empirically determine the circular Gaussian kernel that minimizes the RMS difference between the convolved and interpolated IBIS \halpha{} line widths and the ALMA brightness temperature maps.
The minimization is performed independently at each time in the overlapping set $\{t_O\}$ and then averaged over the time series to produce a final convolution kernel with standard deviation of $1.28\arcsec$, or a FWHM of $3.03\arcsec$.
This is comparable to the ALMA beam size determined by \fnc{CLEAN}.
As a side effect, during the minimization process we found an additional spatial offset of $(1.65\arcsec,0.22\arcsec)$ between the IBIS and ALMA data series that exists in the \citet{Kobelski:2022} data set.
The additional offset has been included in all of the present work.
We stress that a single spatial offset is applied to the entire self-aligned data series, as opposed to a different offset being applied to each time step; the latter would bias our results.
After convolution we interpolate (and therefore down sample) the IBIS data to the same spatial locations as the ALMA pixel centers.
The excellent correlation between the ALMA brightness temperature and the \halpha{} line width, together with the higher spatial resolution and the long-duration and uninterrupted nature of IBIS data series, allow us to study the effects of the spatial and temporal features of the ALMA Band 3 data.

\begin{figure}
    \centering
    \includegraphics[width=0.75\linewidth]{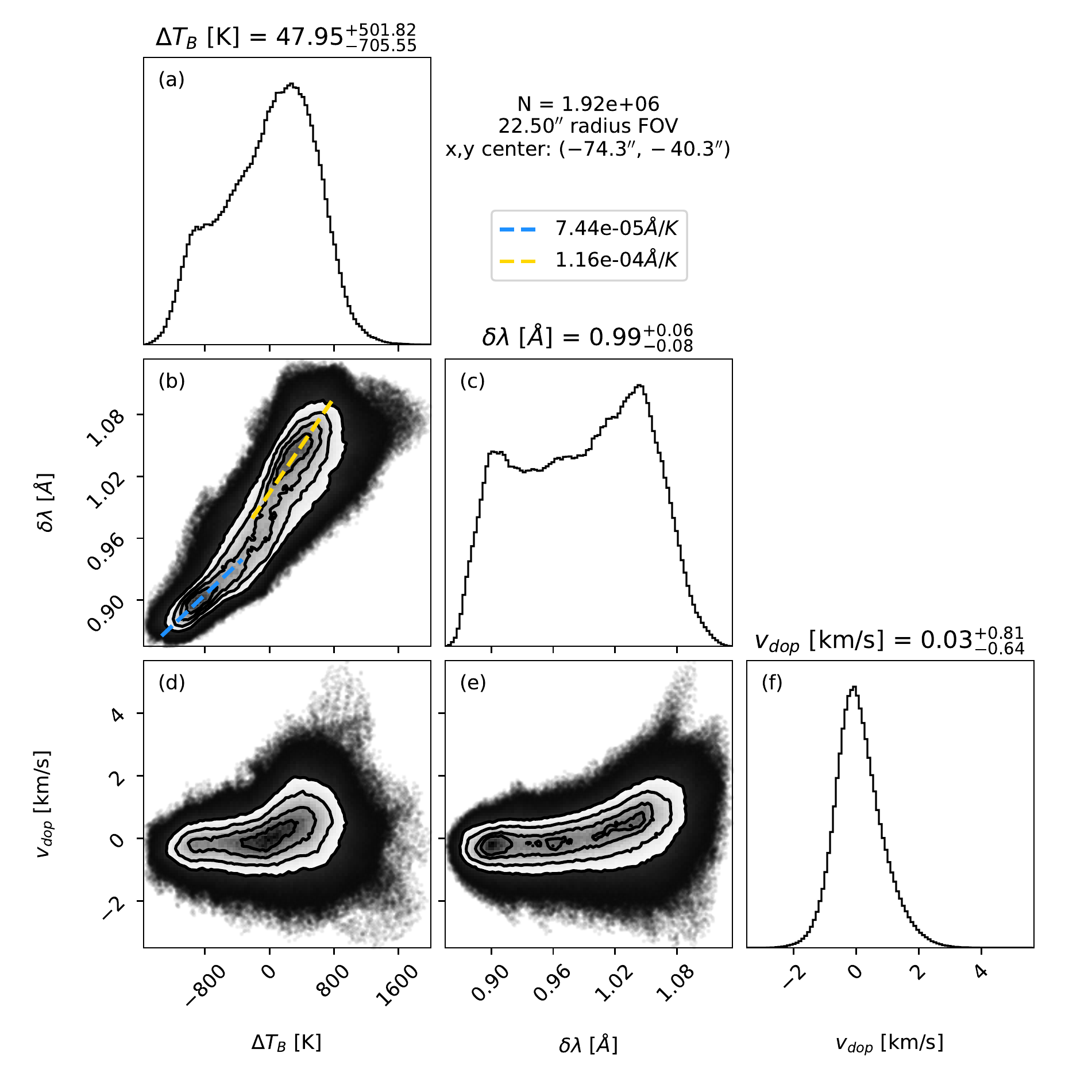}
    \caption{1D distribution for the relative Band 3 brightness temperature (a), \halpha{} line width (c), and \halpha{} Doppler velocity (f), and the 2D joint distributions for $(\Delta T_B,\linew)$ (b), $(\Delta T_B,v_{dop})$ (d), and $(\linew{},v_{dop})$ (e), generated from pixels within a $22.5\arcsec$ radius from the ALMA beam center.  The distributions contain a total of $1.92\times 10^6$ pixels over the 685 joint time steps $\{t_O\}$.  The contours in the joint distributions mark the $20\%$ quintiles, e.g., $20\%$ of the data is between each contour, and the remaining $20\%$ (or $3.8\times10^5$ pixels) are in the outer cloud of individual points.  The blue (yellow) dashed lines show fits to the cores of the cooler (hotter) regions.  The zero point of the ALMA temperature scale is the average over the data and should be close to the value of 7300 K reported by \citet{White:2017}.}
    \label{fig:JDF_alltimes}
\end{figure}

\section{Joint Distribution}\label{sec:joint}
\figref{fig:JDF_alltimes} shows the joint distribution between the relative brightness temperature $\Delta T_B$, the convolved and interpolated \halpha{} line width, and the convolved and interpolated Doppler velocity for all times in the set $\{t_O\}$. 
The figure was created with the \fnc{corner.py} python module \citep{Foreman-Mackey:2016}, which produces a rasterized histogram with contours in equal steps of quintiles ($20\%$ of the data) in the high-density regions of the distribution function, and individual clouds of points in the lowest (outer) regions with the final $20\%$ of the data.
The distribution is generated from all pixels within a 30 pixel (22.5\arcsec{}) radius of the ALMA beam center.
We used this rather conservative definition of the cospatial field of view between the instruments in order to avoid oversampling outer regions of the reconstructed ALMA images where the contrast diminishes, and also avoid potential issues at the edges of the IBIS field of view discussed below.
This reduced field of view still retains enough pixels, $\sim 1.92$ million throughout the overlapping 685 time steps in the set $\{t_O\}$, to give good statistical weight to our results.
Panels (a), (c), and (f) show the 1D distributions of $\Delta T_B$, \linew{}, and $v_{dop}$, respectively.
Panel (b) shows the joint distribution for ($\Delta T_B,\linew{}$), Panel (d) for $(\Delta T_B,v_{dop})$, and Panel (e) for $(\linew,v_{dop})$.

The joint distribution between the \halpha{} line width and the 3 mm brightness temperature in Figure~\ref{fig:JDF_alltimes}(b) shows a bimodal distribution, with a low-$T_B$, low-$\linew$ cluster and a high-$T_B$, high-\linew{} cluster.
The two lobes show different linear trends but have large scatter.
We estimated the linear trend of each lobe using the central moment of each distribution core, defined as the inner $20\%$ of the data.
This definition primarily excluded the broad, high temperature spread of points in the upper right of the distribution, and the resulting fits should be treated as ``by eye.''
The lower temperature, narrower line width region has a shallower slope of $7.4\times10^{-5}$ \AA{}/K (blue line in the figure), and the higher temperature, larger line width region has a steeper slope of $1.2\times10^{-4}$ \AA{}/K (yellow line).
The hotter regions are typically cospatial with the underlying strong magnetic field concentrations (see \figref{fig:overview}).
Looking at the animations of \figref{fig:tblwpddshist} it is clear that the collection of high temperature and broad line width points in the upper right are associated with short-lived dynamic events.
A single linear fit to all the data gives a slope of $9.4\times10^{-5}$ \AA{}/K.

The two lobes in the full joint distribution which were used for the low temperature and high temperature fits in \figref{fig:JDF_alltimes} are barely discernible in the joint distribution for individual times, making instantaneous fits unreliable.
Given the spread in slopes for the high-T, low-T, and full fits, a reasonable estimate for the uncertainty is simply $\pm2\times10^{-5}$\AA{}/K.
The variation of $2\times10^{-5}$ is consistent with the spread in slopes determined using a single linear fit to the joint distribution at individual times (e.g., without trying to separate into hotter and cooler distributions).
On the other hand, the fitted slope at any given individual time has a $95\%$ confidence level of $\approx\pm0.15\times10^{-5}$\AA{}/K, so the time variation of the determined slope has much more variation than allowed by the fit to the slope at any particular time.
The variation in slope seems to be due to a combination of dynamic events and the relative distribution of hotter versus cooler regions in the small, cospatial FOV; see the animation of \figref{fig:overview} and discussion below.

We generated multiple similar distributions by varying the radius that defines the overlapping field of view for the two data series.
As we mentioned in the introduction, ALMA's spatial sensitivity gradually reduces away from the beam center.
This causes a reduction in the contrast of $T_B$ with respect to the center of the field of view, in addition to smoothing of the spatial features.
For the IBIS data, the variable seeing conditions at the DST during our observations caused the field of view to occasionally shift.  
Although we corrected these shifts \textit{post facto} during the self-alignment of the IBIS data \citep[][\S2.5.1]{Kobelski:2022}, they still make the edges of the nominal field of view subject to intermittent data loss\footnote{The solid yellow patch at the bottom of the IBIS FOV in \figref{fig:spatial-psd}(a) is due to one of these intermittent data losses.}.
Using a safely smaller field of view, centered on the ALMA beam center, mitigates both of these issues but limits the range of solar features being sampled and consequently can alter the properties of the measured distribution.

In our case, the ALMA FOV is fairly well centered near the north end of the negative polarity on the western side of the underlying magnetic bipole, as seen in \figref{fig:overview}(a), so the area closest to the beam center is slightly biased towards higher temperatures.
Immediately north of the beam center is the large, somewhat ``W''-shaped region that has the lowest values of temperature, magnetic flux, and line width.  
The regions of more moderate temperature and line width, with reduced physical contrast, are in fact further out from the ALMA beam center, in the region of reduced instrumental sensitivity.
As we increase the area used to generate the distribution, the bimodal lobes in \figref{fig:JDF_alltimes}(b) tend to fill in towards each other.
The same two-slope character remains, but the ``knee'' on the bottom side of the distribution becomes even more prominent at the intersection of the two lobes around (-100 K, 0.95\AA{}).
Note that unlike the ALMA data, there is no \textit{a priori} reason for the IBIS data to trend toward the mean at the edge of the field of view.

In summary, our choice of a $45\arcsec$ diameter circle to define the cospatial FOV, indicated by the red circle in \figref{fig:tblwpddshist}, is our attempt to balance these two competing forms of bias (effects at the edge of the field of view versus uneven sampling of solar features).  
The most important point is that the main features of the joint distribution $(T_B,\linew)$ in \figref{fig:JDF_alltimes}(b) are essentially unchanged regardless of our chosen field of view, even though the individual underlying 1D distributions can change appreciably: for instance, which peak dominates the \halpha{} line width distribution ($0.9$ \AA{} or $1.05$ \AA{}) in panel (c) varies with the FOV as it covers more or less of one of the features.
We thus conclude that the bimodal distribution with two different slopes is a real feature of the data, although we are not certain what physical effect would cause it, or if it could be due to some other unknown systematic error.

The joint distributions involving the Doppler velocity also show a change in behavior for the cooler regions versus the hotter regions, \figref{fig:JDF_alltimes} Panels (d) and (e).
The joint distribution of ALMA with the IBIS Doppler velocity shows essentially no trend below the median ALMA temperature at $\Delta T_B = 0$ and a positive trend above that (Panel d).
This same behavior is found in the joint distribution of the IBIS line width and Doppler velocity using only the IBIS data (Panel e), which again gives us confidence that the trend is physical.
This may be a signature of hot upflows surrounding magnetic concentrations, possibly with dynamic counterparts, but those details will need to be investigated in a subsequent work.

\section{Time series analysis}\label{sec:time}
The IBIS \halpha{} line width data series, with its longer baseline and more continuous coverage, provides a useful reference for understanding the characteristics of the power spectrum of the ALMA 3 mm brightness temperature fluctuations.
The ALMA and IBIS data series each have non-uniform temporal sampling due to the intermittent calibration observations (ALMA) and occasional bad atmospheric conditions (IBIS).
Both constitute missing data in what are otherwise quite regularly sampled time series, but the character of the missing data is very different between the two data series.  
For IBIS, the data have short pauses with essentially random spacing and duration, whereas the calibration sequences for ALMA are extensive (3 minutes) and regular (every 10 minutes).
The latter produces especially strong windowing effects that can significantly affect the interpretation of power spectra generated from the data \citep{VanderPlas:2018}.
We therefore used the method of Lomb \citep{Lomb:1976} and Scargle \citep{Scargle:1982} as implemented in the \fnc{astropy.LombScargle} package and described in \citep{VanderPlas:2012,VanderPlas:2015} to estimate the power spectral density (PSD) of each data series.
We calculated the power spectra using the same range of frequencies and spectral bins for all data series, but have verified that this produced identical power spectra to the case of letting the software package algorithmically determine the optimal set of bins.

The \fnc{astropy} package allows for several different normalizations of the PSD.  We used the default normalization which, for a time series $s(t)$ sampled at $N$ times $\{t_n\}$, calculates the power spectrum normalized to the total power of the mean subtracted time series:
\begin{gather}
    \fnc{PSD}(f) = \frac{\chi_\text{ref}^2 -\chi^2(f)}{\chi_\text{ref}^2},\label{eq:psd}\\
    \text{where}\qquad \chi_\text{ref}^2 = \sum_n\bigl\langle (s_n - \langle s\rangle)^2\bigl\rangle = \sum_n\langle s_n^2\rangle - \langle s\rangle^2\label{eq:totalpower}
\end{gather}
is the mean squared deviation, $s_n = s(t_n)$,  and $\chi^2(f)$ is the goodness-of-fit of the least-squares fit of a sinusoidal model to the measured data at a given frequency, i.e.
\begin{gather}
    \chi^2(f) = \sum_n\Bigl[s_n - \bigl[a(f) \sin2\pi f (t_n-\phi_f) + b(f) \cos2\pi f (t_n-\phi_f)\bigr]\Bigr]^2.
\end{gather}

The standard Lomb-Scargle power spectrum given in Equation \eqref{eq:psd} is normalized to the total power in a measured (and mean-subtracted) time series, $\chi_\text{ref}^2=\langle s^2 \rangle$ which makes it a measure of the \emph{relative} distribution of power over frequency for a given time series (a single spatial location in the $\Delta T_B$, $\linew{}$, or $v_\text{dop}$ data series).
At a given frequency $0\leq PSD(f)\leq 1$, where 0 would mean that the best-fit model has zero amplitude ($a = b = 0$), and 1 would mean that the entire time series is perfectly fit by a single frequency sinusoid $f$.
For evenly sampled data, the Lomb-Scargle spectrum is related to the Fourier spectrum by
\begin{gather}
    \fnc{PSD}_F(f_i) = \frac{1}{2}\chi^2_\text{ref}\fnc{PSD}(f_i)
\end{gather}
where $f_i = \frac{i}{T}$ are the typical Fourier frequencies.
In this section Figure \ref{fig:total-power-map} displays total power in physical units and all the remaining figures use the arbitrary Lomb-Scargle normalized units (ADU).

In Section~\ref{sec:avgpowermaps}, we calculate the spatially averaged power spectra for each data series and compare how temporally downsampling the sets $\{t_A\}$ and $\{t_I\}$ to the set $\{t_O\}$ affects the average power spectra.
This analysis provides evidence that at least some of the previously reported lack of oscillatory power in the 3-5 min period range for the 3 mm observations \citep{Jafarzadeh:2021} may be an artifact of the calibration sequences.

We then consider the spatial distribution of oscillatory power computed for each of data series in Section~\ref{sec:avgpowermaps}.
Here we see the spatial patterns typically found in other, similar chromospheric data sets.
By using temporal and spatial downsampling of the \halpha{} we confirm that the spatial variations in power the ALMA brightness temperature fluctuations follow those for the \halpha{} line width, which provides further evidence that the calibration windows are masking the detection of \pmode{} power in the ALMA data.

\subsection{Spatially Averaged Power Spectra}\label{sec:avgpowermaps}
\begin{figure}
    \centering
    \includegraphics[width=0.95\textwidth]{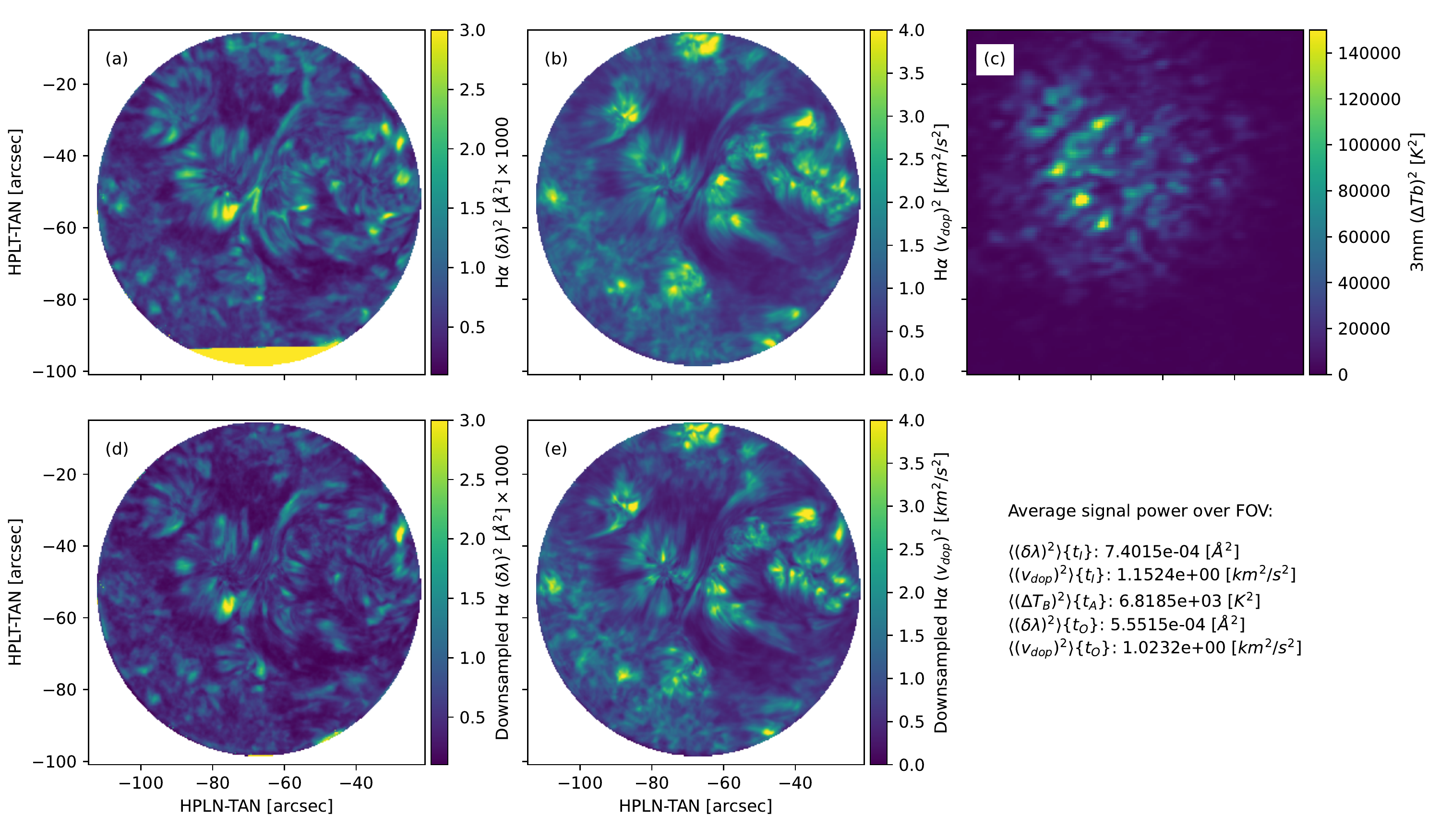}
    \caption{Total time-averaged power maps for different data series: (a) \linew{}, (b) $v_\text{dop}$, (c) $\Delta T_B$, each calculated over the respective time series $\{t_I\}$ and $\{t_A\}$.  Panels (d) and (e) repeat the first two panels for the IBIS data, but use the down sampled, overlap time series $\{t_O\}$.  The spatially averaged total power is given for each data series in the lower right panel.}
    \label{fig:total-power-map}
\end{figure}

\figref{fig:total-power-map} shows spatial maps of the temporally averaged total fluctuation power for the various data sets in physical units, e.g, $\chi_\text{ref}^2(x,y)$ from Equation \eqref{eq:totalpower} evaluated at spatial location $(x,y)$.
Panel (a) shows the power in the line width fluctuations, (b) in the Doppler velocity, and (c) in the brightness temperature, calculated using the full temporal set for each variable ($\{t_I\}$ for IBIS and $\{t_A\}$ for ALMA).
Panels (d) and (e) show the average power for the line width and Doppler data using the reduced time series $\{t_O\}$.
The spatial average over the FOV for each map is given in the lower right panel of the figure; RMS variations can be obtained by taking the square root of the provided values.

The primary effect of temporally down sampling the IBIS data from $\{t_I\}$ to $\{t_O\}$ is the introduction of the $\approx 3$ min ALMA calibration windows into the IBIS time series.
Comparing panels (a) and (d) to (b) and (e), this down sampling appears to have a greater effect on the calculation of power in fluctuations of the line width than for the Doppler velocity.
This is borne out by comparing the spatially averaged power in each case, which is reduced by $25\%$ for the former but only $11\%$ for the latter.
The power in the ALMA brightness temperature fluctuations bears some resemblance to that of the line width data, particularly the ``W'' shaped suppressed power near the center of network cell around $(-75\arcsec, -30\arcsec)$, and some of the brightest kernels in the ALMA data, but the correspondence is not particularly striking.

We next calculated the power spectral density for each of the data series described in Sections \ref{sec:obs} and \ref{sec:dataprep}, i.e., the full resolution ALMA and IBIS data series as well as the temporally and/or spatially down sampled versions of each.
We are not concerned with the joint distribution here, so the individual field of view for each instrument was used, as opposed to the $45\arcsec$ common field of view considered in Section \ref{sec:joint}.
As described by Equation \eqref{eq:psd}, the power spectrum of a single signal (i.e., a single spatial location) is normalized to its total power.
The power spectra at different spatial locations can then be ensemble averaged to determine the relative distribution of power across all frequencies.

\begin{figure}
    \centering
    \includegraphics[width=0.75\textwidth]{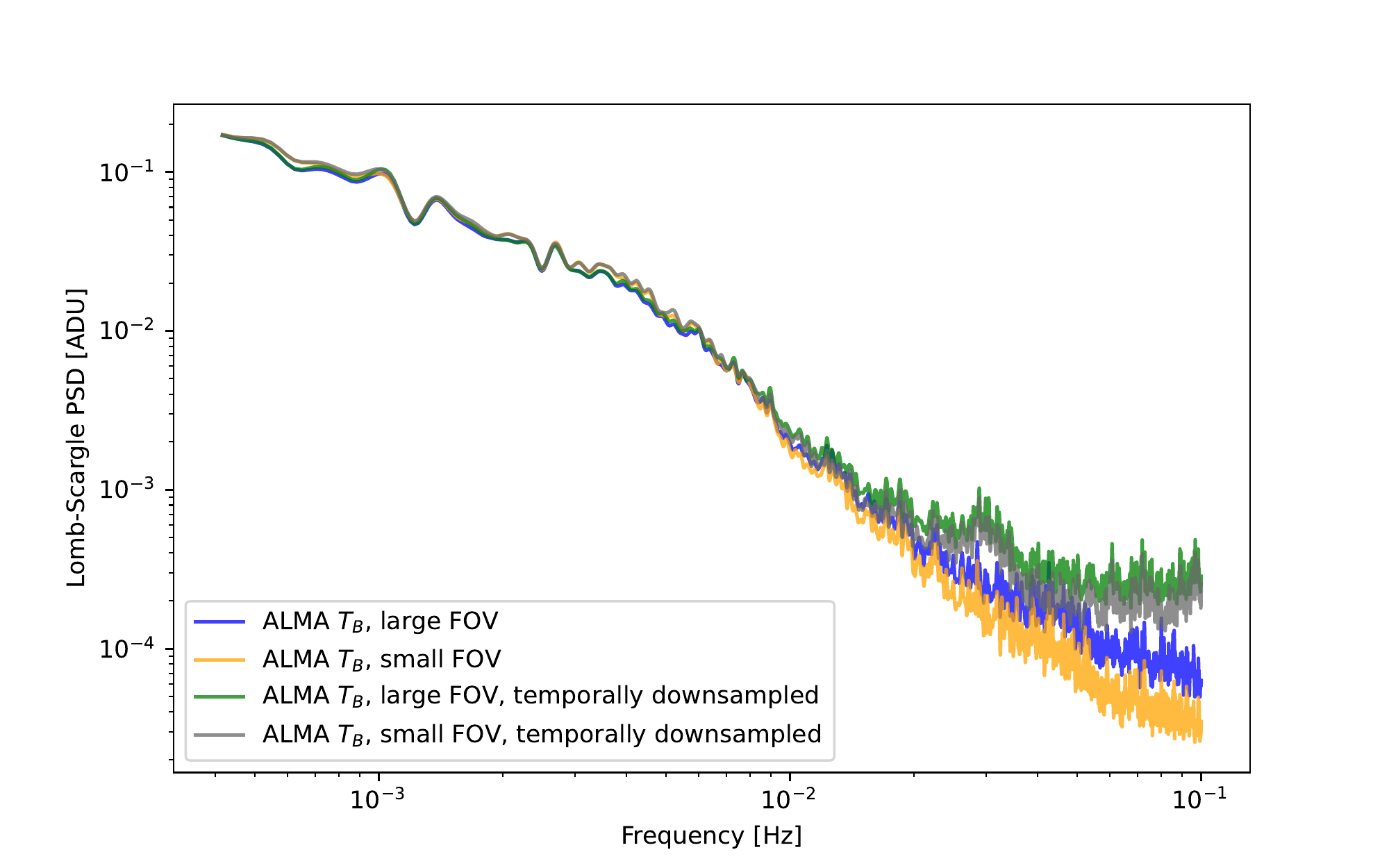}
    \caption{Spatially averaged power spectra for different sub-samples of data series.  ALMA temperature, large $67\arcsec$ FOV (blue); ALMA temperature, small $45\arcsec$ FOV (orange); ALMA temperature large FOV, $\{t_O\}$ (green); ALMA temperature small FOV, $\{t_O\}$ (grey).}
    \label{fig:spatial-psd-alma}
\end{figure}

\figref{fig:spatial-psd-alma} shows the spatially averaged power spectral density of the ALMA 3 mm brightness temperature using four different ways of spatially or temporally downsampling the data. 
The blue curve shows the power spectrum computed using the full temporal cadence $\{t_A\}$ and a $67\arcsec$ diameter FOV about the beam center.
The orange curve is the PSD for the full temporal cadence averaged over a $45\arcsec$ FOV.
The green curve is the PSD calculated with the reduced cadence overlapping time series $\{t_O\}$ and averaged over the $67\arcsec$ FOV.
Finally, the gray curve uses the overlapping set $\{t_O\}$ and smaller FOV.

The power spectra are essentially identical for all cases below about $10\mhz$ and show an apparent spectral break around $5\mhz$, from a shallower to steeper slope.
The larger FOV has slightly more power relative to the smaller FOV at high frequencies, above $\sim20\mhz$.
The temporal down sampling has a more dramatic effect, increasing the white-noise floor above about $40\mhz$ and producing a conspicuous bump at $30\mhz$.
A similar bump at $30\mhz$ appears to be present in Figure 5(d) of \citet{Chai:2022}, which shows power spectra for ALMA in quiet Sun regions.
Finally, we note that the strong deviations from the trend at low frequencies, near $1.5\mhz$ and $2.5\mhz$, are artifacts of the time series window function (see \citet{VanderPlas:2018}).

\begin{figure}
    \centering
    \includegraphics[width=0.75\textwidth]{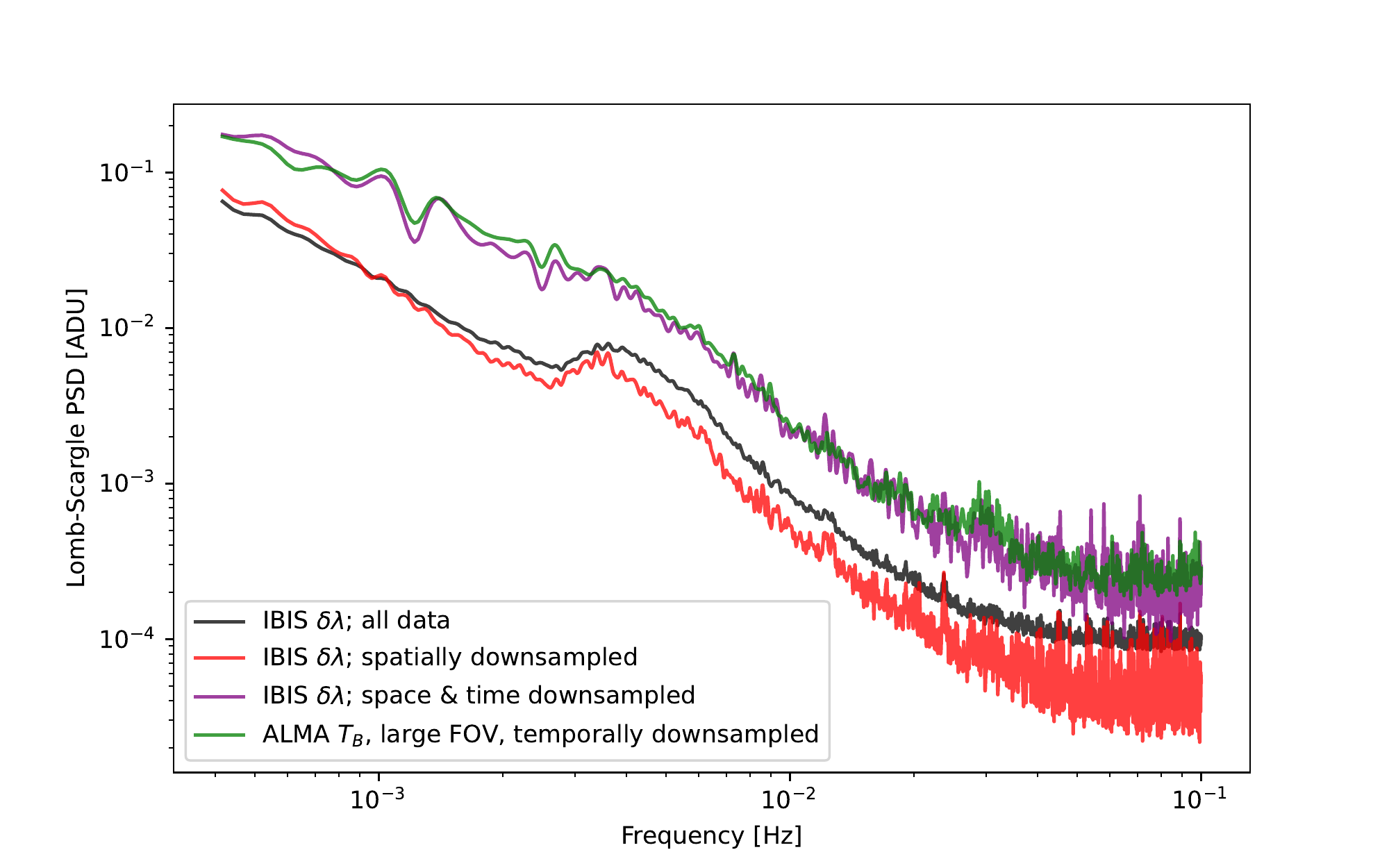}
    \caption{Spatially averaged power spectra for different sub-samples of data series.  All IBIS line width data (black); spatially down sampled IBIS line width (red); spatially and temporally down sampled IBIS line width (purple); ALMA temporally down sampled temperature  $67\arcsec$ FOV (green).}
    \label{fig:spatial-psd}
\end{figure}

In \figref{fig:spatial-psd} we repeat the analysis of \figref{fig:spatial-psd-alma} for the IBIS \halpha{} line width data with the original resolution and several down sampled versions.
The power spectrum for the temporal down sampled ALMA data from Figure \ref{fig:spatial-psd-alma} (green curve) is included for context.
The black curve is calculated using the IBIS data at full spatial and temporal resolution, the red curve uses the convolved and spatially down sampled IBIS data, and the translucent purple curve uses the spatially and temporally down sampled IBIS data, i.e., IBIS sampled at the ALMA spatial resolution and in the set $\{t_O\}$.
Here we see that spatially down sampling has little effect on the power derived from the line width data while temporally down sampling the IBIS data to include the ALMA calibration windows results in power spectral density curves that lie nearly on top of each other.
To quantify the above statement, the Pearson correlation coefficient between the temporally-and-spatially down sampled IBIS data and the temporally down sampled ALMA data, i.e., what should be the two most similar data sets, is $0.9895$.  For comparison, the correlation coefficient for the full resolution version for each curve (blue line from Figure \ref{fig:spatial-psd-alma} to black line from Figure \ref{fig:spatial-psd}) is 0.9731, and the full ALMA series (blue) to spatially reduced IBIS (red) is 0.9464.

The relative offset between the full temporal IBIS power curve (black or red) and the temporally down sampled power (purple) is due to the standard Lomb-Scargle normalization, which does not include a factor for the number of sampled times $N$. 
If present, this factor would differ between the two time series by the ratio of the elements in the set $\{t_I\}$ versus $\{t_O\}$, (that is, 1787 to 685, or roughly 2.6:1), and be uniformly applied over all frequencies, thus producing a constant offset in the log-scaled plot.
Because we are currently analyzing only the \emph{relative} distribution of power over frequency (e.g., the shape of each curve), we opt not to perform any additional normalization when plotting our results in this section.
Thus, the difference in number of sampled times from the underlying signal manifests as a constant offset in the log-scaled plots while, at the same time, preserving their shape. 
In the current context, this actually helps to distinguish between the various power curves: the power spectra resulting from a single continuous signal's time series with two different samplings, but both being well sampled, will be shifted by the ratio of the sampling, while changes in the shape of the power spectra will be due to under-sampling or windowing effects.

\begin{figure}
    \centering
    \includegraphics[width=0.75\linewidth]{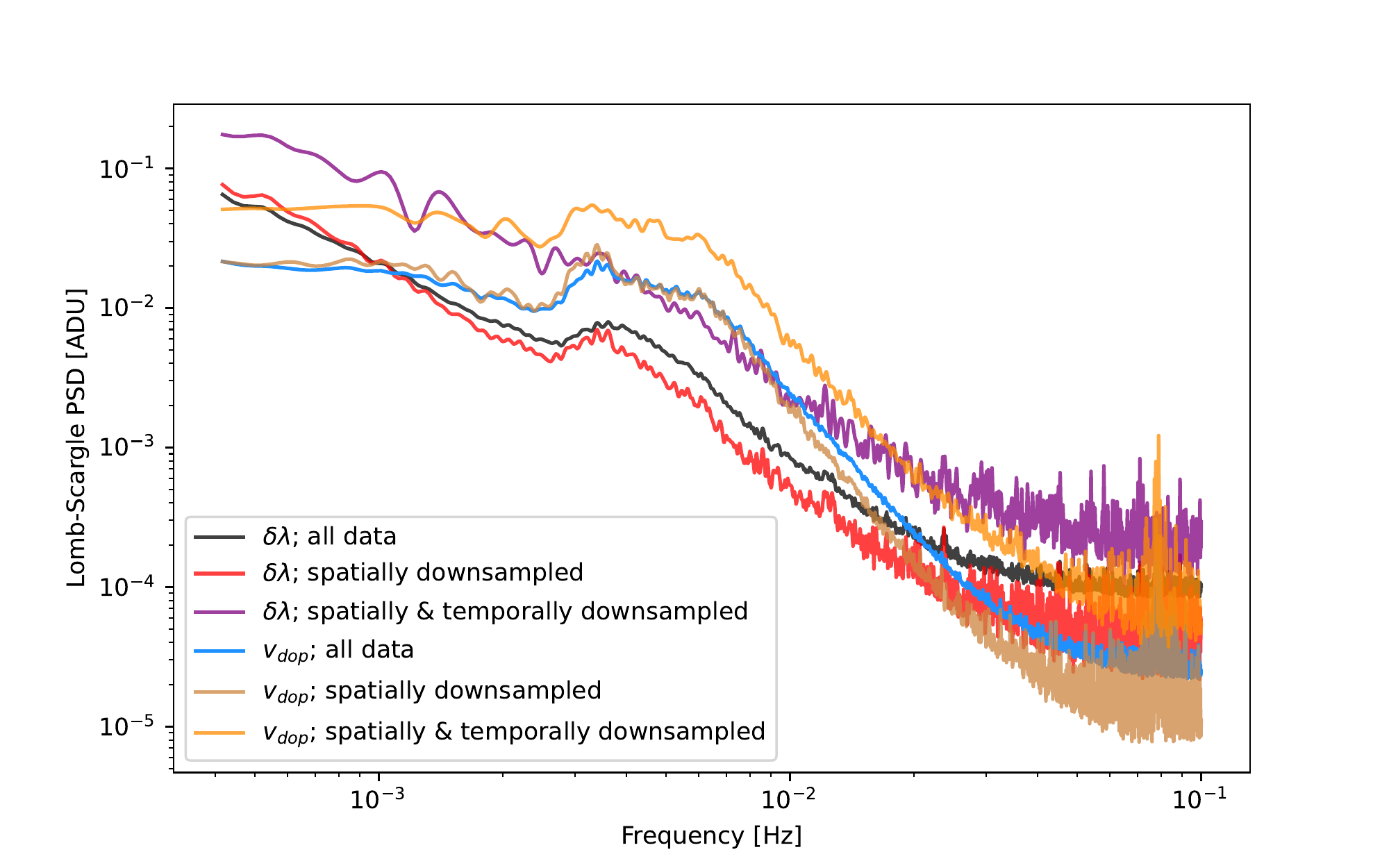}
    \caption{Spatially averaged power spectra of the IBIS Doppler velocity time series averaged over the full IBIS field of view for different spatial and temporal sampling schemes. The spectrum derived from the full data set is shown in blue, spatially down sampled in brown (which nearly overlaps the blue line below 10\mhz{}), and spatially-temporally down sampled in orange.  The line width power spectra from Figure \ref{fig:spatial-psd} are included for reference.}
    \label{fig:average-ibis-full}
\end{figure}

To place each curve on a roughly appropriate physical power scale the relative curves in Figures \ref{fig:spatial-psd-alma}-\ref{fig:average-ibis-full} should be divided by the appropriate number of timesteps included in the data set as well as the average power indicated in \figref{fig:total-power-map}.
However, we caution that these temporally unevenly sampled datasets, each of which include some level of spatial coherence that are folded into the spatial average, make a precise calculation of the absolute power a more involved analysis than we are attempting here.

The prominent bump at $4\mhz$ in the black curve in \figref{fig:spatial-psd} marks the solar \pmode{} frequency band in the full resolution \halpha{} line width data.
That band contains considerable power across the typical 3 and 5 min oscillation windows.
The effect of either spatially or temporally down sampling the data is immediately apparent by comparing the translucent red or purple curves to the black curve.
Spatial down sampling the IBIS data does not eliminate the \pmode{} signal (red), while temporally down sampling the IBIS data does (purple).
The temporal down sampling does produce the vertical shift in the line by the ratio of the number of sampled times, as discussed above, while the change in shape is due to undersampling the signal with the new temporal sampling. 
The temporal down sampling produces a power spectrum that is closely matches the ALMA PSD up to $10\mhz$ (compare the purple curve to the green curve) and remains well correlated at higher frequencies.
Given that the temporal down sampling of ALMA data from 2 to 4 $\unit{sec}$ cadence produced very little change in the resulting power spectrum below $\sim10\mhz$, this analysis provides strong evidence that the apparent lack of \pmode{}s in the ALMA data is a spectral effect due to the interval and duration of the calibration sequences.

As a final comparison, we calculate the spatially averaged power spectra, including spatially and temporally down sampled versions, of the IBIS \halpha{} Doppler velocity data.
The results are shown in \figref{fig:average-ibis-full} where we also include the curves for the \halpha{} line width for comparison, using the same colors as in \figref{fig:spatial-psd}.
The power spectrum for the full resolution Doppler data is shown in blue, spatially downsampled in light brown, and both spatially and temporally downsampled in orange.
The Doppler velocity power spectra shows markedly different behavior than those for the line width (black, red, purple) or brightness temperature (not shown, but the purple line here is largely the same).
The \pmode{} bump is again prominent in the Doppler data but with a different distribution compared to that in the line width.
The temporal down sampling does not appear to suppress the \pmode{}s in the Doppler velocity data as it did for the line width data and, presumably, the ALMA data.
However, this may be due to the flat slope of the Doppler power distribution below 2\mhz{} in the original data combined with the strong apparent oscillations from the windowing function at 1.5 and 2.5\mhz.

\subsection{Frequency Integrated Maps of Power Spectral Density}\label{sec:powermaps}
As mentioned in the introduction, there have been conflicting reports as to how the spatial distribution of power in different frequency bands varies across the ALMA Band 3 field of view.
Some authors find substantial variations \citep{Molnar:2021,Chai:2022}, others do not \citep{Patsourakos:2020, Narang:2022} or find somewhat ambiguous results \citep{Jafarzadeh:2021}.
We find ourselves in the former camp, with seemingly well resolved spatial variations in oscillatory power across the field of view.
As we will see, the variations in the ALMA brightness temperature align well with those derived from the \halpha{} line width once the two data series were spatially and temporally sampled on the same grids.

\begin{figure}
    \centering
    \includegraphics[width=\textwidth]{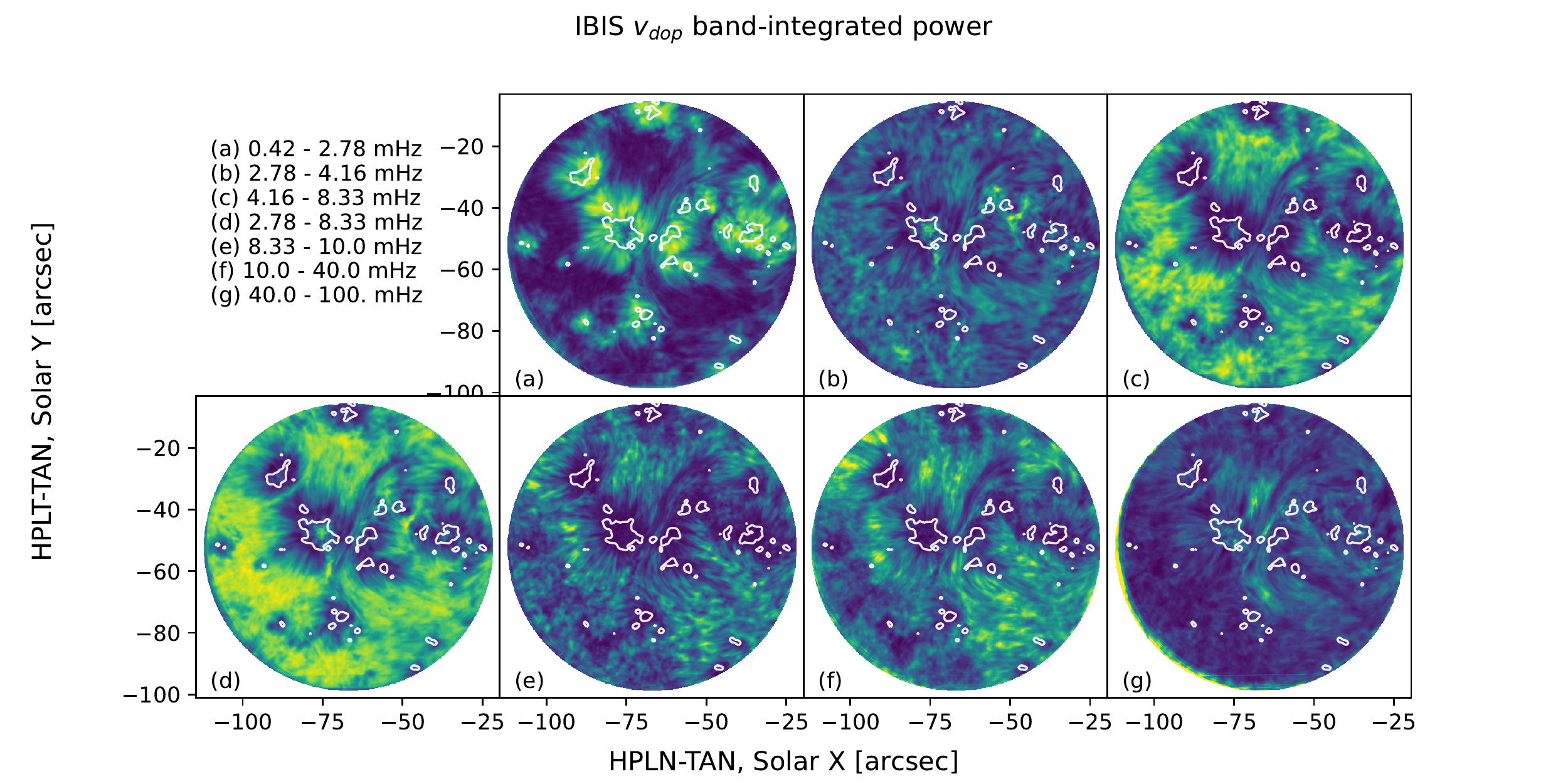}
    \caption{Spatial maps of the frequency-integrated power spectral density of the IBIS \halpha{} Doppler velocity over the seven frequency bands indicated in the upper left.  These bands correspond to: (a) low frequencies; (b) 5 min oscillations; (c) 3 minute oscillations; (d) combined \pmode{}s; (e) near-cutoff; (f) mid frequencies; (g) high frequencies.  The spatial coordinates correspond to the solar position at the beginning of the ALMA observations at 2017-03-21 15:42:13 UT.  Note that the color scale is adjusted separately for each panel as the data span $\approx40 \unit{dB}$.  Higher power is bright yellow, lower power is dark blue.  The white lines are the 50 G contours of unsigned LOS magnetic flux from HMI.}
    \label{fig:IBIS-dp-psd-map_full}
\end{figure}

\begin{figure}
    \centering
    \includegraphics[width=\textwidth]{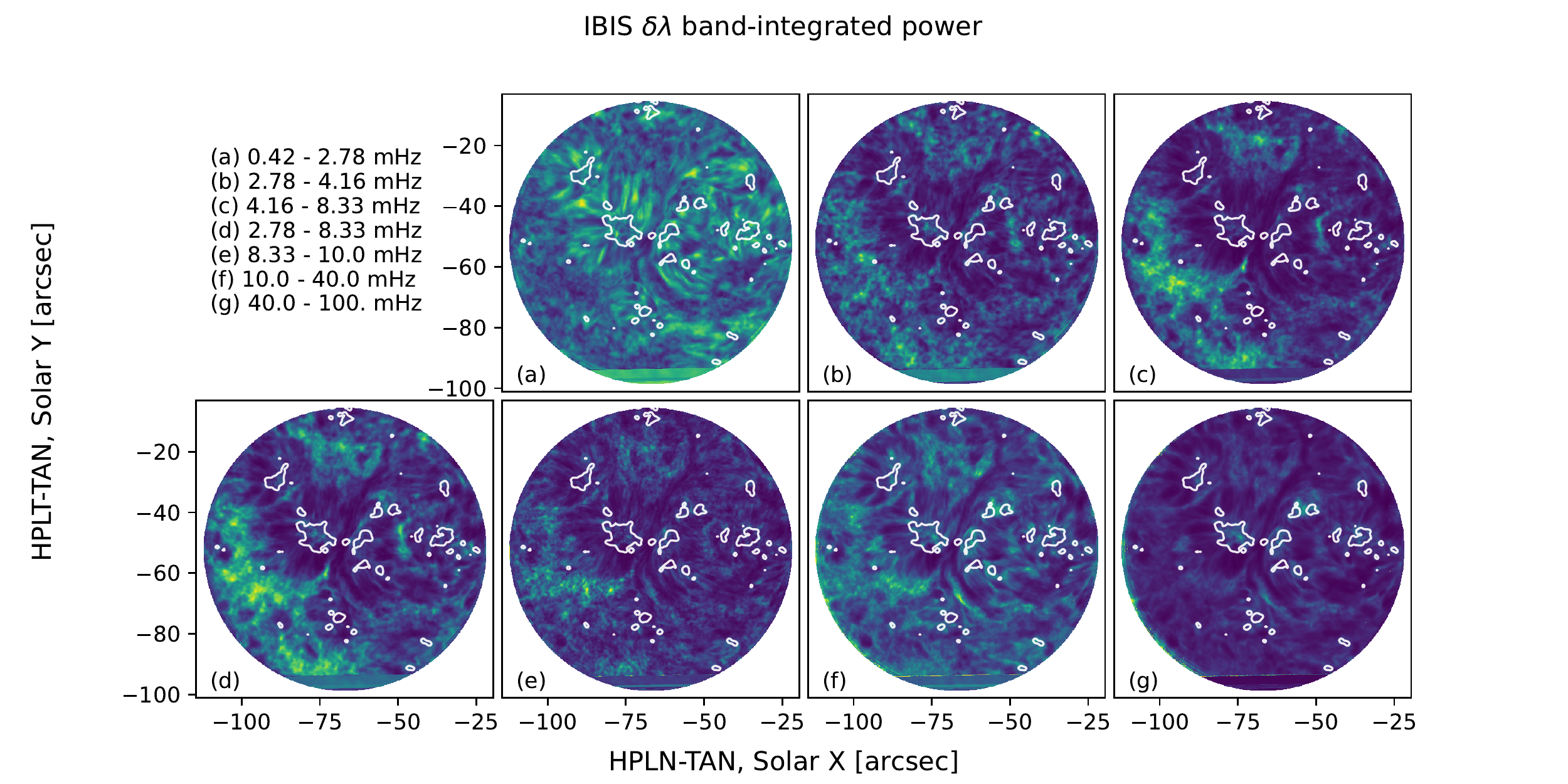}
    \caption{The same as \figref{fig:IBIS-dp-psd-map_full} but for the full temporal and spatial resolution IBIS \halpha{} line width.}
    \label{fig:IBIS-lw-psd-map_full}
\end{figure}

\begin{figure}
    \centering
    \includegraphics[width=\textwidth]{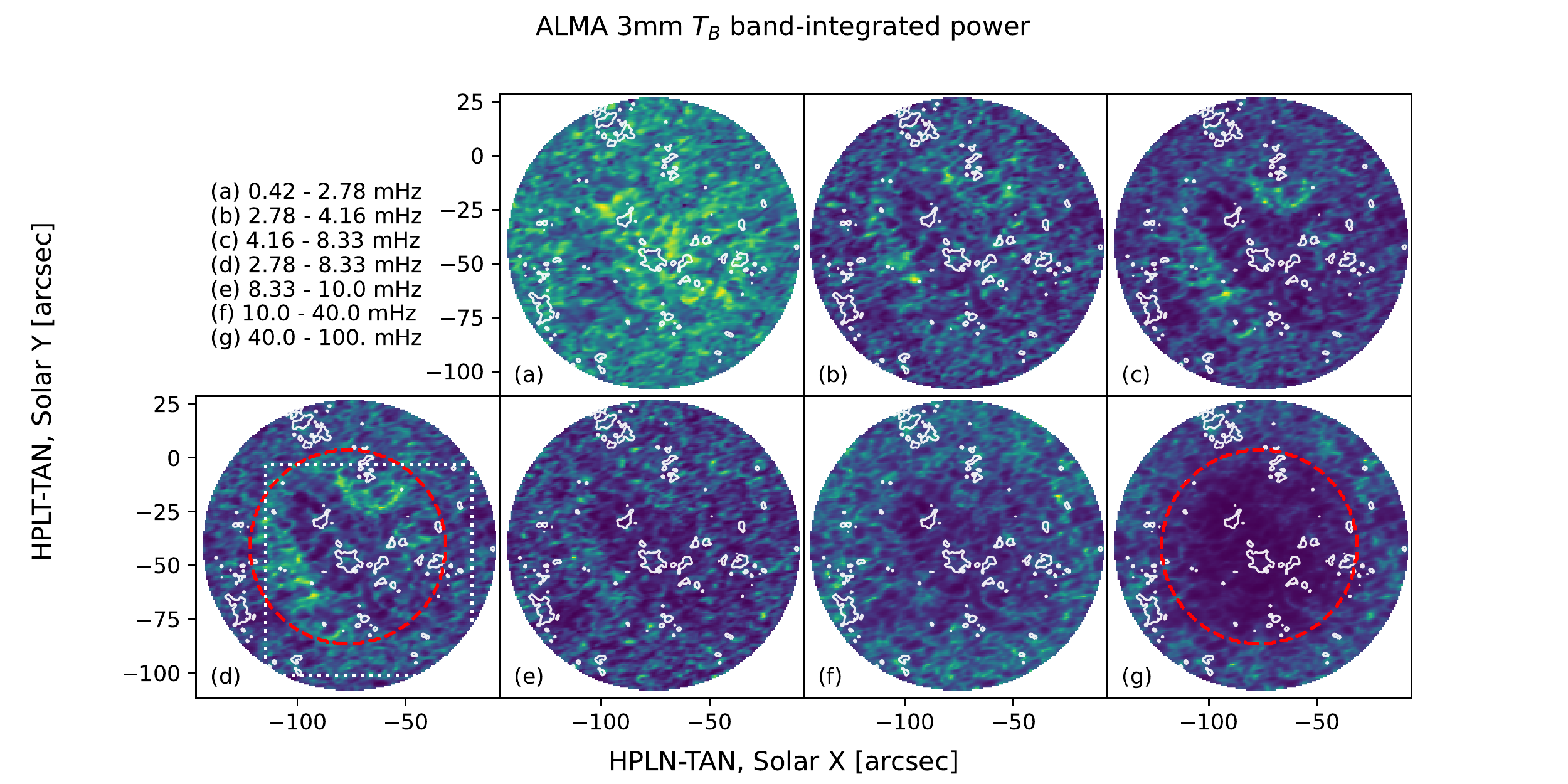}
    \caption{Same as \figref{fig:IBIS-dp-psd-map_full} but for the ALMA 3 mm brightness temperature and a slightly larger field of view.  The outer boundary has a $67\arcsec$ radius while the red dashed circle shows a $45\arcsec$ diameter, both relative to the ALMA beam center.  The white dotted rectangle in Panel (d) indicates the IBIS FOV.}
    \label{fig:alma-psd-map-full}
\end{figure}

\begin{figure}
    \centering
    \includegraphics[width=\textwidth]{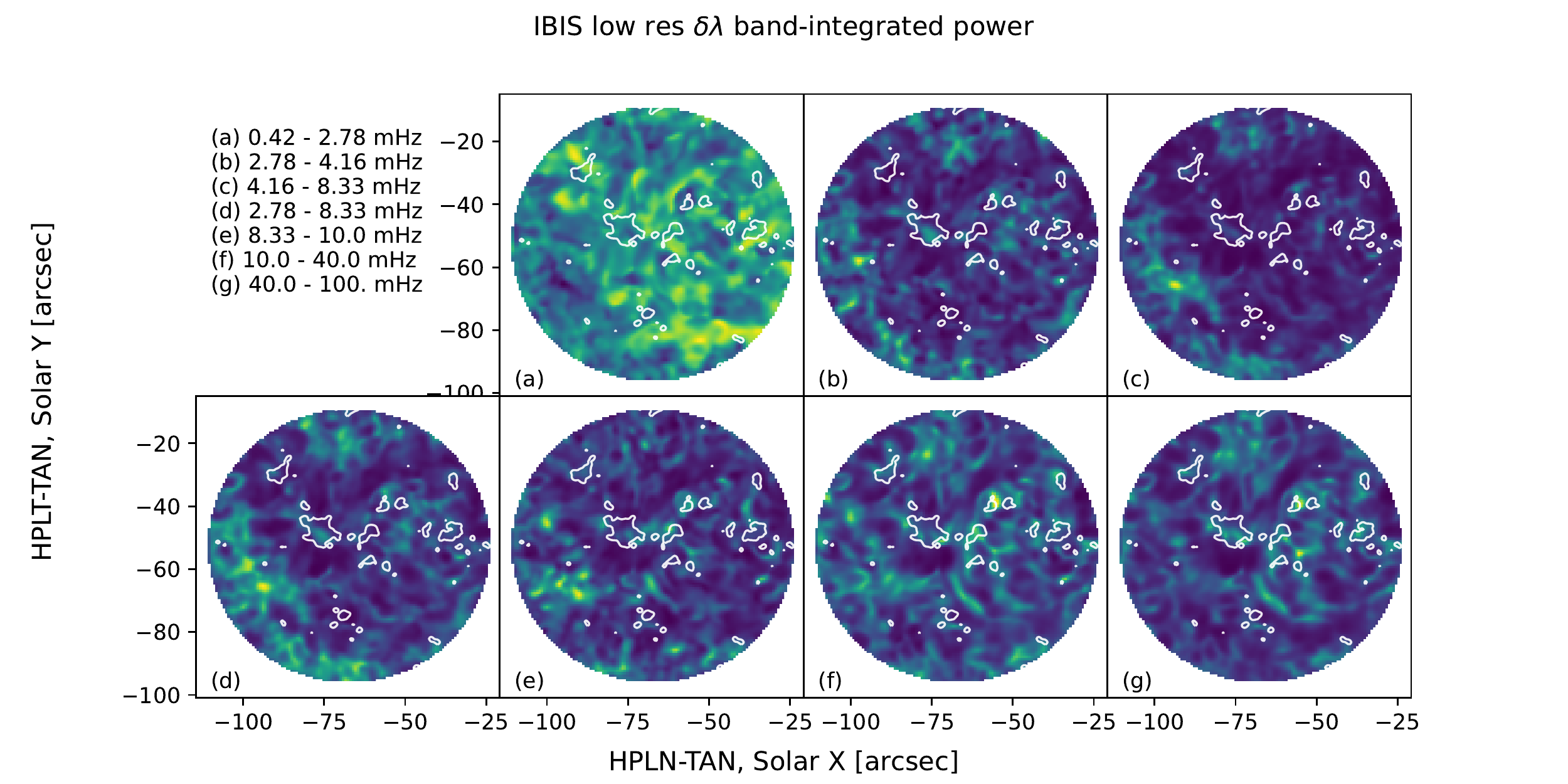}
    \caption{Same as \figref{fig:alma-psd-map-full} but for the IBIS line width data spatially convolved and downsampled to match ALMA, and temporally subsampled to the overlapping times $\{t_O\}$.}
    \label{fig:IBIS-lw-psd-map_downsample}
\end{figure}

In Figures \ref{fig:IBIS-dp-psd-map_full} to \ref{fig:IBIS-lw-psd-map_downsample} we show the spatial distribution of power in different temporal frequency bands for each of the data series.
\figref{fig:IBIS-dp-psd-map_full} is generated using the full spatio-temporal IBIS \halpha{} Doppler velocity, \figref{fig:IBIS-lw-psd-map_full} the full spatio-temporal IBIS \halpha{} line width, \figref{fig:alma-psd-map-full} the ALMA Band 3 brightness temperature, and \figref{fig:IBIS-lw-psd-map_downsample} the IBIS \halpha{} line width after spatial convolution and interpolation to the ALMA resolution and pixel locations.
The sub-panels in each figure show the frequency-integrated power in the ranges: (a) 0.42-2.78\mhz, (b) 2.78-4.16\mhz, (c) 4.16-8.33\mhz, (d) 2.78-8.33\mhz, (e) 8.33-10\mhz, (f) 10-40\mhz, and (g) 40-100\mhz.
Phenomenologically, these bands correspond to (a) low frequencies, (b) the 5 min \pmode{}s, (c) the 3 min \pmode{}s, (d) the total \pmode{} range, and (e,f,g) three higher frequency bands that attempt to capture the transition from evanescent waves just above the acoustic cutoff (if one can reasonably be defined) to more freely propagating MHD waves in regimes where the WKB\footnote{See \citet{Weinberg:1962} for a complete description of applying Wentzle-Kramers-Brillouin approximation to the MHD equations.} approximation is expected to hold, e.g., where the characteristic wavelength of a disturbance is smaller than the typical gradient scale of the medium.

Comparing the Doppler velocity power in different frequency bands in \figref{fig:IBIS-dp-psd-map_full} makes it immediately apparent how the different dynamics correspond to different features on the Sun.  
The distribution of power in the low frequency band shown in Panel (a) is essentially the inverse of that found in the 5\mhz{} (c) and higher bands.
The areas of highest power at low frequency lie above the stronger network magnetic flux concentrations that can be seen in HMI or Hinode/SP observations (see \figref{fig:overview} and \citet{Kobelski:2022} Figure 2) but cover a much larger physical area as the magnetic field expands to form the magnetic canopy in the chromosphere.
These same regions have suppressed power at higher frequencies, at least up to $40\mhz$ in panel (f), though this trend has essentially disappeared above that frequency, as seen in panel (g).
These features are well known and define the so-called magnetic shadows and acoustic halos \citep{Brown:1992,Braun:1992, Judge:2001, Vecchio:2007,Vecchio:2009}.
We note that the largest, strongest region does show a compact suppression of power at its center near $(-75\arcsec,-50\arcsec)$ in the lowest frequency bin, as seen in panel (a), but this behavior in the other magnetic patches is less conclusive.

The distribution of power in the line width data, \figref{fig:IBIS-lw-psd-map_full}, shows a much different pattern.  
Here the most prominent concentrations of power are seen in the 5\mhz{} band in the quiet Sun regions to the southeast of the central magnetic bipole.
This is also the region with rather moderate line widths, not the large line width regions surrounding the network flux patches, nor the the very narrow width region immediately north of the central bipole (\figref{fig:tblwpddshist}).
The strong power in the quiet Sun regions is likely due to acoustic shocks \citep{Vecchio:2009}.

The final two figures in this section show the spatial power maps for the ALMA 3 mm brightness temperature, in \figref{fig:alma-psd-map-full}, and the convolved, downsampled IBIS \halpha{} line width, in \figref{fig:IBIS-lw-psd-map_downsample}.
According to our results in \secref{sec:joint} and \secref{sec:avgpowermaps}, these two figures should be very similar, and indeed they are.
The ALMA data in Panel (a) at low frequencies show a slight enhancement in the central portion of the field of view relative to the surrounding area. 
This pattern is consistent with the full resolution IBIS \halpha{} line width data in \figref{fig:IBIS-lw-psd-map_full}(a) and the interpolated data in \figref{fig:IBIS-lw-psd-map_downsample}(a).
In contrast, the V-shaped (or heart-shaped) central area of the ALMA field of view shows lesser power than the surrounding areas in all the frequency bands above 3\mhz{}, panels (b-f).
Again, this same pattern is visible in the PSD maps of the sub-sampled IBIS data in bands up to 10\mhz, \figref{fig:IBIS-lw-psd-map_downsample} (b-e), but becomes less apparent the two higher frequency bands.
This behavior is consistent with the spatially averaged spectra in \figref{fig:spatial-psd-alma}, where the ALMA and IBIS data series match well up to around 10\mhz, then diverge (compare the purple curve to either the blue or orange curve).

At the highest temporal frequencies we consider in \figref{fig:alma-psd-map-full} (g), the ALMA data shows a strong radial pattern that appears dominated by the sensitivity of the spatial reconstruction more than anything else.
Still, some of the features continue to line up between the ALMA and IBIS PSD maps, such as the compact high power point near $(-50\arcsec,-40\arcsec)$ that is discernable in panels (f) and (g) of Figures \ref{fig:IBIS-lw-psd-map_full}, \ref{fig:alma-psd-map-full}, and \ref{fig:IBIS-lw-psd-map_downsample}.

\section{Discussion}\label{sec:discussion}

We have confirmed the result of \citet{Molnar:2019} that the 3 mm brightness temperature and \halpha{} line width are strongly correlated.
We find a bimodal distribution in the 2D histogram of these variables, corresponding to hotter network regions and cooler internetwork regions.
The \halpha{} line width increases more steeply with respect to the ALMA brightness temperature in hotter regions than in the cooler regions.  
The slope that we find is different than \citet{Molnar:2019}, and some change in slope might be due to the different definition of the \halpha{} line width from the IBIS data.
The slope and clustering may depend on the variety of observed solar features in the two sets of observations. 
Our modest network patch, located very close to disk center, has a greatly reduced range of observed ALMA temperatures compared to the somewhat stronger network patch near an active region studied in \citet{Molnar:2019} and \citet{Molnar:2021}.
Understanding the underlying reason for the significant variation in the range of observed temperatures in Band 3, even for relatively similar solar features, will require further investigation.

Our use of the intensity at $\pm$0.75\AA{} to set the upper ``continuum'' threshold introduces crosstalk in the \halpha{} line width measurement when the line is sufficiently broad to suppress the wing intensity at those wavelengths. This appears to cause the metric to saturate at a maximum line width of around 1.2\AA{}, but also an increasing reduction of the expected line width values for widths above 1.0\AA{}.
The saturation is clearly evident when reprocessing the \citet{Molnar:2019} data (from April 2017) to more closely match the data and line-fitting parameters in the current analysis.
However, the greatly reduced temperature range of the March 2017 data series appears to stay mostly in the linear regime.
This makes a direct comparison between the two data sets difficult to interpret.
In general, using the wing intensity at $\pm$1.0\AA{} or greater in the calculation of the line width is generally preferred as it avoids this saturation effect.

What is certain, despite the disparities just discussed, is that the ALMA Band 3 brightness temperature and \halpha{} are strongly correlated diagnostics.
That good correlation provided us with an independent reference to help separate systematic effects in the ALMA data from observed solar behavior in the subsequent power spectral analysis.
In particular, we have attempted to address the confusing melange of reports of oscillatory power in the ALMA Band 3 data.
In summary, we found:
\begin{itemize}
    \item When averaged over space, the \pmode{}s are only clearly visible in the IBIS \halpha{} data, and show up clearly in the power spectra of both \linew{} and $v_{dop}$.
    \item Spatially down sampling IBIS \linew{} data to ALMA spatial resolution does not remove \pmode{} signature.
    \item Temporally down sampling IBIS \linew{} data does remove the obvious \pmode{} signature.
    \item The preceding steps make the spatially averaged \linew{} and the $T_B$ power spectra match extremely well, especially below 10$\mhz$.
    \item Both the spatially averaged ALMA PSD curve and the spatially resolved power maps show little difference for the full versus temporally overlapping subsets below $15\unit{mHz}$; e.g., temporally downsampling from $\sim2$ to $\sim 4$ seconds does not significantly alter the ALMA power spectra.
    \item The spatial maps of ALMA oscillatory power match the maps of spatially and temporally downsampled \halpha{} line width power.
    \item A conspicuous bump at $30\unit{mHz}$ is present in the temporally downsampled ALMA data, but in no other data series in our data set.  It does seem to be present in the average quiet Sun PSD reported in Figure 5d of \citet{Chai:2022}, and in the network flux near an active region reported in \citet{Molnar:2021}.
\end{itemize}

In \secref{sec:avgpowermaps} we showed that for our data series the lack of significant power in the \pmode{} bands in the ALMA data is consistent with it being an artifact of the temporal sampling, particularly the duration and spacing of the calibration windows.
This conclusion is based on the transition from a clear \pmode{} signal in the full \halpha{} \linew{} power spectrum, as calculated over the entire time series, to a lack of such signature after applying the ALMA calibration windows to the \halpha{} data and then recomputing the spatially averaged power spectrum.
The latter now closely matches the ALMA spectrum, as shown in \figref{fig:spatial-psd}.
This result likely holds for other data series as well.
Our conclusion required the combination of the 3 mm brightness temperature and the more continuous \halpha{} line width data.

It is clear from \figref{fig:IBIS-lw-psd-map_downsample}(b-d) that the strongest concentrations of \pmode{} power in the IBIS \halpha{} \linew{} data are in the quiet Sun just outside the network flux regions, following the typical magnetic shadow/acoustic halo pattern \citep{Braun:1992,Brown:1992,Rajaguru:2013}.
While less clear in the lower resolution ALMA data, we find that the spatial variation of \pmode{} power is still present.
\citet{Loukitcheva:2004} predicted strong \pmode{} signals at sub-mm wavelengths in non-magnetic atmospheres.
This was seemingly confirmed by observations with the BIMA \citep{Loukitcheva:2006}.
With the preceding arguments, we find that it holds in our data series as well. 
Significantly, we find spatial distributions in the 3mm power, and these vary with frequency band in tandem with the line width power maps.
Contrarily, \citet{Patsourakos:2020} reported no such spatial variation in their band-integrated power maps (see discussion immediately preceding their \S2.1).

\citet{Chai:2022} found unambiguous detection of 3 min oscillations in sunspot umbra, but not the surrounding penumbra or quiet Sun in the ALMA data.
In contrast, cotemporal \halpha{} observations also displayed clear 5 min oscillations in the penumbra, but unfortunately did not extend into the quiet Sun.
The ALMA observations had 3.6 minute calibration scans in between 10.25 minute blocks of on-target observations, similar to our own, which suggests that the lack of a prominent \pmode{} signature in the quiet Sun could be due to the calibration sequences.
However, this could also be influenced by proximity to the sunspot, whose magnetic canopy has a much larger extent compared to our humble bipolar network patch.

\citet{Jafarzadeh:2021} provide extensive examples of Lomb-Scargle power spectral density similar to what we have presented in this work.
Their examples cover both ALMA Band 3 and Band 6 observations and a wide variety of solar targets. 
Their power spectra display a similar knee, or apparent spectral break, to that seen in Figure \ref{fig:spatial-psd-alma} with no obvious peak at the \pmode{} frequencies for 8 out of the 10 datasets they analyzed (see their Figure 11).
They attribute the lack of clear \pmode{} signals to the spatial distribution of strong magnetic elements in and around the FOV in the majority of their datasets:
``While we discussed above various possible scenarios to explain these oscillatory behaviours, we conjecture that the lack of 3-min (5.5 mHz) oscillations may be a result of (a) the ‘umbrella’ effect due to the magnetic canopy, (b) power suppression in the presence of strong magnetic fields, (c) significant variations in the height of formation, or (d) waves not displaying temperature fluctuations at these frequencies.''

We do find some enhanced \pmode{} power in our ALMA data in the quiet regions just outside of magnetic network concentrations, as seen in \figref{fig:alma-psd-map-full}(d).
The magnetic field in the present data set is not markedly different from some of the data sets for which \citet{Jafarzadeh:2021} reported a lack of \pmode{} oscillations.
The higher resolution and complementary view provided by the map of PSD derived from the \halpha{} \linew{} in \figref{fig:IBIS-lw-psd-map_full}(d) shows more clearly the familiar pattern of power shadows and halos centered on the magnetic concentrations \citep{Braun:1992,Brown:1992,Judge:2001,Vecchio:2007,Vecchio:2009,Rajaguru:2013}.
Comparing Figures \ref{fig:IBIS-dp-psd-map_full}-\ref{fig:alma-psd-map-full}, the exact pattern of shadows and halos depends on the frequency band of the map and physical properties of the diagnostic, e.g., oscillations in velocity versus intensity, and formation height.
In the \pmode{} band for all diagnostics we find some power at the very center of the strong magnetic patches, surrounded by the lower power shadow, and the strongest power in the quiet regions surrounding the shadows.
This is most evident for the strongest magnetic element in our FOV, the negative footpoint of the central bipole at $(-75,-50\arcsec{})$.
In terms of oscillatory power, our small bipolar region acts like a miniature active region, despite the fact that the magnetic concentrations are not strong or large enough to produce even a micropore.
Again, this is consistent with previous work \citep{Vecchio:2007}.
In contrast, \citet{Patsourakos:2020} did not find evidence for shadows or halos in their band-integrated power maps, and \citet{Narang:2022} also found little spatial variation in the maps of oscillatory power across the field of view in their Band 6 observations targeting a region of plage.
While we have not studied Band 6 data, given our findings for the Band 3 data, it is possible that the calibration sequences have affected their results.

\section{Conclusions}\label{sec:conclusion}
We presented results that combine optical and millimeter wavelength diagnostics using several of the publicly available data series described in \citet{Kobelski:2022} and accessible at \url{https://share.nso.edu/shared/dkist/ltarr/kolsch/}.
By undertaking a joint analysis of two related diagnostics, namely the ALMA Band 3 brightness temperature and the IBIS \halpha{} line width, we found evidence that the previously reported lack of observed \pmode{}s in at least some of the ALMA Band 3 data may be an artifact of the temporal sampling.
While we cannot conclusively demonstrate that \pmode{}s are present in our ALMA Band 3 observations, our combined findings of their clear presence in the \halpha{} data, the strong correspondence in the spatially-averaged \halpha{} and ALMA PSDs, and the strong correspondence between the two diagnostics in band-integrated spatial maps are all highly suggestive.
This result provides strong motivation for adjusting the operational model for solar observations with ALMA from what was done in Cycle-4.  
Calibration windows could be shorter and/or less frequent, albeit with an impact on the reliability of the phase corrections.
Alternatively, semi-random spacing of calibration windows would reduce temporal windowing artifacts.
We also find that spatially resolved maps of oscillatory power in Band 3 data integrated over temporal frequency bands do reveal suppression of power near magnetic concentrations and (slight) enhancements outside of those suppression regions.
Given the weakness magnetic field, it is not yet obvious what magnetic configuration is associated with the enhancements.

The spatial pattern in the power maps of the ALMA data correspond well to maps of power generated from the spatially downsampled \halpha{} line width data.
However, the variance in the spatial distribution of power is much lower in the downsampled \halpha{} data compared to the original data.
Given the strong correlation between the line width and brightness temperature and the nearly equivalent spatially-averaged power spectra of the two data sets once placed on the same spatial and temporal grids, we believe that our identification of magnetic shadows and power halos around the network concentrations in the ALMA data is correct.

\section*{Conflict of Interest Statement}
The authors declare that the research was conducted in the absence of any commercial or financial relationships that could be construed as a potential conflict of interest.

\section*{Author Contributions}
Funding for this project was secured by A.R.K. (principal investigator) with  L.A.T. and S.A.J.
The majority of the data analysis was carried out by L.A.T., with supporting data analysis by S.A.J., K.R., A.R.K., M.M., and G.C.
The manuscript was prepared by L.A.T. with supporting work from S.A.J. and A.R.K. 
All authors participated in discussion of the analysis and were responsible for reading and editing the manuscript.

\section*{Funding}
The work of L.A.T., A.R.K., and S.A.J. was supported in part by NASA Heliophysics Supporting Research grant 80NSSC19K0118.
L.A.T., S.A.J., M.M, G.C., and K.R. were supported by the National Solar Observatory (NSO).  NSO is managed by the Association of Universities for Research in Astronomy, Inc., and is funded by the National Science Foundation.
M.M. was supported for part of this work by a FINESST fellowship with grant number 80NSSC20K1505 and by the DKIST Ambassador Program, funding for which is provided by the National Solar Observatory, a facility of the National Science Foundation, operated under Cooperative Support Agreement number AST-1400405.
K.P.R. also acknowledges the support of NASA under the grant 80NSSC20K1282.

\section*{Acknowledgments}
The authors thank Nicholas Luber for their help in previous work calibrating the ALMA data used here.

This paper makes use of the following ALMA data: ADS/JAO.ALMA$\#$2016.1.00788.S. ALMA is a partnership of ESO (representing its member states), NSF (USA) and NINS (Japan), together with NRC (Canada), MOST and ASIAA (Taiwan), and KASI (Republic of Korea), in cooperation with the Republic of Chile. The Joint ALMA Observatory is operated by ESO, AUI/NRAO and NAOJ.
Data in this publication were obtained with the Dunn Solar Telescope of the National Solar Observatory, which is operated by the Association of Universities for Research in Astronomy, Inc., under cooperative agreement with the National Science Foundation.  
This research has made use of NASA’s Astrophysics Data System Bibliographic Services.
Analysis was carried out primarily in the Python programming language and made use of the AstroPy \citep{Astropy:2013,Astropy:2018} and SunPy \citep{Sunpy:2020} packages.
Figures were prepared using Matplotlib \citep{Hunter:2007}.

\section*{Data Availability Statement}
Publicly available datasets were analyzed in this study. This data can be found here: \url{https://share.nso.edu/shared/dkist/ltarr/kolsch/}.

\bibliographystyle{frontiersinHLTHnFPHY} 
\bibliography{almaibis}

\end{document}